\documentclass[sigconf]{aamas} 

\usepackage[ruled,vlined]{algorithm2e}
\usepackage{svg}
\usepackage{balance} % for balancing columns on the final page

\makeatletter
\gdef\@copyrightpermission{
  \begin{minipage}{0.2\columnwidth}
   \href{https://creativecommons.org/licenses/by/4.0/}{\includegraphics[width=0.90\textwidth]{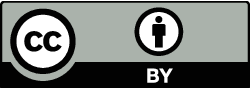}}
  \end{minipage}\hfill
  \begin{minipage}{0.8\columnwidth}
   \href{https://creativecommons.org/licenses/by/4.0/}{This work is licensed under a Creative Commons Attribution International 4.0 License.}
  \end{minipage}
  \vspace{5pt}
}
\makeatother

\setcopyright{none}
\acmConference{}{}{}
\acmYear{}
\acmDOI{}
\acmISBN{}
\renewcommand\footnotetextcopyrightpermission[1]{%
  \textit{Preprint. Under review.}}

\acmSubmissionID{<<2766>>}

\title[]{Gossip Agents: Reputation through Opinion Dynamics over Evolving Networks}

\author{Yutong Wu}
\affiliation{
  \institution{Rice University}
  \city{Houston}
  \country{United States}}
\email{yw180@rice.edu}

\author{Arlei Silva}
\affiliation{
  \institution{Rice University}
  \city{Houston}
  \country{United States}}
\email{arlei@rice.edu}

\begin{abstract}
Sustained cooperation among self-interested learning agents is essential in many shared environments. Reputation systems are among the best-studied mechanisms for promoting such cooperation. Existing multi-agent reinforcement learning approaches, however, typically assume that reputation is either globally available to all agents or derived solely from an agent's own observations, leaving largely unexplored how reputational information propagates between agents. We study gossip, the exchange of information about third parties, as a mechanism for reputation diffusion over dynamic communication networks. Agents form private opinions from limited observations and exchange them with their neighbors following DeGroot opinion dynamics, while the communication network changes as agents interact and move. A tunable broadcast radius controls the reach of information, spanning private reputation, local gossip, and globally shared reputation. We study this framework in two \(n\)-player social dilemmas with distinct network dynamics: a Prisoner's Dilemma with randomly resampled interaction groups and Harvest, where communication neighborhoods emerge from agents' movements. Our results show that the reach of reputational information shapes cooperation in both environments. Moreover, in Harvest, reputation diffusion increases environmental rewards and leads to more peaceful and sustainable behavior.

\end{abstract}

\newcommand{\BibTeX}{\rm B\kern-.05em{\sc i\kern-.025em b}\kern-.08em\TeX}

\begin{document}

%%% The following commands remove the headers in your paper. For final 
%%% papers, these will be inserted during the pagination process.

%\pagestyle{fancy}
%\fancyhead{}

%%% The next command prints the information defined in the preamble.

\maketitle 

%%%%%%%%%%%%%%%%%%%%%%%%%%%%%%%%%%%%%%%%%%%%%%%%%%%%%%%%%%%%%%%%%%%%%%%%
\section{Introduction}

%Cooperation among self-interested agents is a key factor in deploying learning systems in shared environments and is studied across multiple disciplines, including evolutionary biology~\cite{nowak_98,nowak_06,milinski_02} and multi-agent systems (MAS)~\cite{Wang_22, Oroojlooy_22}. For example, autonomous vehicles negotiate road space with other vehicles and with human drivers; energy and bandwidth markets allocate a finite resource among participants who each prefer more of it. In each case, the interaction is mixed-motive rather than cooperative or zero-sum: agents' objectives are neither fully aligned nor strictly opposed, and behavior that is rational for the individual aggregates into an outcome that is worse for everyone, including the individual~\cite{dietz_03,hardin_68,ostrom_90,ostrom_99}. This is the structure of a social dilemma, and independent learners in multi-agent reinforcement learning (MARL) optimizing their own discounted returns converge reliably to the defecting equilibrium in such settings. The question of what mechanisms are sufficient to sustain cooperation among learning agents is therefore not a peripheral concern about prosociality, but a precondition for deploying populations of such agents at all~\cite{Pennisi_05,fatima_24}.

Cooperation among self-interested agents is central to deploying learning systems in shared environments and has been studied across disciplines, including evolutionary biology~\cite{nowak_98,nowak_06,milinski_02} and multi-agent systems (MAS)~\cite{Wang_22,Oroojlooy_22}. For example, autonomous vehicles share road space with other vehicles and human drivers, while energy and bandwidth markets allocate finite resources among participants with competing demands. Such interactions are often mixed-motive: agents' objectives are neither fully aligned nor strictly opposed, and individually rational behavior can lead to outcomes that are worse for everyone~\cite{dietz_03,hardin_68,ostrom_90,ostrom_99}. This tension between individual and collective interests characterizes a social dilemma, in which self-interested learning can favor defection over sustained cooperation. Understanding which mechanisms can sustain cooperation among learning agents is therefore a fundamental challenge for deploying populations of autonomous agents in shared environments~\cite{Pennisi_05,fatima_24}.

%Reputation is among the oldest and best-studied mechanisms to promote cooperation. Under indirect reciprocity, an agent's past conduct is summarised into a social standing that others consult before deciding how to treat it, so that defection carries a cost imposed not by the victim's retaliation but by the community's subsequent withdrawal of cooperation~\cite{murase_23,vilone_16,giardini_14}. MARL has adopted this idea in several forms, from reputations assigned by a third-party observer under a fixed social norm~\cite{anastassacos_21,orzan_24} to evaluation policies that agents learn for themselves~\cite{LR2}, and learning agents, like human groups, sustain cooperation only when one another's conduct can be tracked over time~\cite{mckee_23}.

Reputation is among the best-studied mechanisms for promoting cooperation. Under indirect reciprocity, an agent's past conduct informs a social standing that others use to determine how to interact with it, allowing defection to carry costs through the subsequent withdrawal of cooperation by others rather than through direct retaliation~\cite{murase_23,vilone_16,giardini_14}. MARL has incorporated this idea in several forms, ranging from reputations assigned by third-party observers according to fixed social norms~\cite{anastassacos_21,orzan_24} to learned evaluation policies~\cite{LR2}. These approaches demonstrate how tracking agents' behavior over time can promote cooperation among agents~\cite{mckee_23}.

In these approaches, however, reputation either reaches every agent at once or remains confined to the agent that observes it, leaving largely unexplored how reputational information can spread through second-hand communication~\cite{sommerfeld_07,feinberg_14}. When a single observer or a fixed norm maintains one score per agent that is available to all participating agents~\cite{anastassacos_21,orzan_24}, reputation is globally available by design. Conversely, when reputation is formed from first-hand interaction~\cite{LR2}, information remains private to the observer, limiting the reputational consequences of behavior that others do not witness~\cite{pan_24,wu_16,molho_21,cruz_21}. Gossip sits between these extremes: agents exchange assessments of third parties, allowing information acquired through direct observation to reach agents who did not witness the underlying behavior. 

Gossip introduces two challenges for reputation systems. First, observers are not interchangeable~\cite{LR2,Hilbe_18,Kawakatsu_24}. Agents form judgments from limited and potentially different information, including partial views of the environment, distinct interactions, and private histories. As a result, agents may hold conflicting opinions, and propagating these opinions may lead to disagreement or even consensus around an inaccurate assessment. Second, the communication network itself may not be fixed. Who can inform whom changes as agents interact~\cite{Pacheco_06,Rand_11,Su_23}. Existing gossip-based approaches, however, typically assume a fixed communication graph~\cite{cooper}.

To address these challenges, we study how reputational information propagates among learning agents that form different opinions from limited observations and communicate over networks that change over time. We introduce a reputation framework that separates three processes: private assessment, reputation aggregation, and information diffusion. First, each agent forms private opinions of the agents it directly observes according to a fixed assessment rule. Second, these private opinions are aggregated into a reputation signal that shapes each agent's reward. Third, agents exchange opinions with their neighbors following the DeGroot model of opinion dynamics~\cite{DeGroot_74}, allowing information acquired through direct observation to propagate to agents that did not witness the underlying behavior. A tunable broadcast radius $\rho$ controls the reach of this information, spanning private reputation, local gossip, and globally shared reputation. The communication network is redrawn at every step, either through the resampling of interaction partners or as a consequence of agents' movement. To the best of our knowledge, this is the first work to study how reputation diffusion through opinion dynamics shapes cooperation among reinforcement learning agents over dynamic communication networks.

Our contributions are as follows:
\begin{itemize}
    %\item We propose a reputation framework that separates private assessment, reputation aggregation, and gossip, and implements gossip as DeGroot averaging over a dynamic network. 
    \item We introduce a decentralized reputation framework that explicitly models how reputational information propagates among learning agents. The framework separates private assessment, aggregation, and gossip, using DeGroot opinion dynamics to exchange private opinions and a tunable broadcast radius to span private, local, and global reputation.
    %\item We apply the framework in two $n$-player social dilemmas: a Prisoner's Dilemma in which agents are regrouped into random tables every round, and the Harvest Sequential Social Dilemma~\cite{leibo_17_multi}, in which neighborhoods follow agents' positions and are therefore shaped by their learned behavior.
    \item  We couple reputation diffusion with dynamic communication networks and study two distinct forms of network evolution: exogenous changes through random regrouping in an $n$-player Prisoner's Dilemma, and endogenous changes induced by agents' learned movement in Harvest~\cite{leibo_17_multi}.
    %\item We provide a detailed analysis of agent behavior in both games and show that a wider broadcast radius improves cooperation in each, though in different ways. Beyond environmental reward, we evaluate Harvest based on social outcome metrics and find that agents using our framework behave more peacefully and sustainably.
    \item We characterize how the reach of reputational information shapes cooperation and collective outcomes. In both games, it encourages cooperation, and in Harvest, it additionally encourages more peaceful and sustainable behavior.
\end{itemize}

\section{Related Work}

%\textbf{Cooperation in Mixed-Motive Games.} Among matrix games, the Prisoner's Dilemma is the canonical model for the conflict between individual and collective interest, and many approaches have been developed to encourage cooperation among participating agents. Opponent-shaping methods let an agent influence how its opponent learns~\cite{Foerster_18, lu_22}, a second line of methods builds reciprocity into the policy itself~\cite{lerer_18,peysakhovich_18}, and a third line introduce partner selection and study how that influences the learner~\cite{anastassacos_21,baker_20}.

\textbf{Cooperation in Mixed-Motive Games.} In matrix games, the Prisoner's Dilemma is the canonical model of the tension between individual and collective interests, motivating a broad range of mechanisms for promoting cooperation among self-interested agents. Opponent-shaping methods promote cooperation by influencing how other agents learn~\cite{Foerster_18,lu_22}, while reciprocity-based approaches incorporate responses to an opponent's past behavior directly into the agent's policy~\cite{lerer_18,peysakhovich_18}. Another class of methods introduces partner selection, allowing agents to choose whom to interact with and thereby creating incentives for cooperative behavior~\cite{anastassacos_21,baker_20}.

Sequential social dilemmas extend matrix games to Markov environments, where cooperation and defection correspond to temporally extended behaviors rather than individual actions~\cite{leibo_17_multi}. P\'erolat et al.~\cite{Leibo_17} instantiate this framework in common-pool resource (CPR) appropriation and introduce Harvest, which serves as our second testbed. Subsequent work has developed several mechanisms for promoting cooperation in Harvest. One class augments agents' rewards with social preferences, including inequity aversion~\cite{hughes_18} and altruism~\cite{mckee_20}. McKee et al.~\cite{mckee_20}, for example, show that introducing altruistic agents can increase collective return, but at the cost of equality, as altruistic agents effectively subsidize the consumption of their self-interested peers. A second class of studies allows agents to directly influence one another's incentives through mechanisms such as gifting~\cite{Lupu_20} and binding contracts~\cite{christoffersen_23}. While these approaches promote cooperation by modifying agents' incentives or preferences, our work instead studies how cooperation emerges when information about agents' behavior propagates through reputation and gossip. %A central question of our work is whether cooperation can arise from a reputation system without prosocial preferences being built into the agents, and what this means for peacefulness, sustainability, and equality.

\textbf{Reputation in Multi-Agent Reinforcement Learning.} A growing body of work incorporates reputation into MARL as a mechanism for promoting cooperation. Existing approaches differ primarily in how reputation is constructed and how reputational information is made available to other agents. One class assigns each agent a reputation, often represented as a discrete label, that is maintained by a third-party observer or centralized mechanism and made available to other agents~\cite{anastassacos_21,orzan_24,mckee_23}. In these approaches, reputation is public by construction: each agent is associated with a shared reputation value rather than potentially different assessments held by different observers. Consequently, the propagation of reputational information is not explicitly modeled; agents do not need to exchange their assessments, and access to reputation does not depend on the communication network. In our work, by contrast, agents form private opinions from limited observations, and reputational information propagates through communication with neighboring agents before being aggregated into reputation.

%Closest to our work are LR2~\cite{LR2} and COOPER~\cite{cooper}, in which agents hold private assessments of one another. In LR2, however, each agent plays against a fixed set of neighbors, and reputation assessments never travels beyond that neighborhood. In COOPER, agents play against randomly chosen opponents and share their assessments over a social network, but they are trained only on two-player games and communicate over a static graph. Our work addresses both limitations: we study $n$-player games in which opponents change from round to round and reputation travels by gossip over a network that evolves over time.

Closest to our work are LR2~\cite{LR2} and COOPER~\cite{cooper}, which similarly allow agents to maintain private assessments of one another. In LR2, however, agents interact within fixed neighborhoods, and their assessments do not propagate beyond the agents they directly encounter. COOPER allows agents to interact with randomly selected opponents and exchange assessments through a social network, but considers two-player games and assumes a static communication graph. Our work extends these settings to \(n\)-player social dilemmas in which both interaction partners and communication neighborhoods can change over time. We consider both exogenous network dynamics, where agents are randomly regrouped between rounds, and endogenous network dynamics, where communication neighborhoods emerge from agents' learned movement.

%\textbf{Evolutionary Biology.} Reputation has long been studied in evolutionary biology as the basis of indirect reciprocity~\cite{nowak_98,nowak_06,milinski_02}, with further work on which assessment rules sustain cooperation~\cite{ohtsuki_06} and on the design of reputation systems~\cite{dellarocas_03}. Much of this theory assumes that reputations are common knowledge. When that assumption is dropped and individuals hold private views, disagreement undermines cooperation~\cite{Hilbe_18}, and no cooperative norm is evolutionarily stable when individual opinions are statistically independent~\cite{murase_24}. Gossip is the standard remedy: it can substitute for direct observation~\cite{sommerfeld_07,feinberg_14}, and Kawakatsu et al.~\cite{Kawakatsu_24} derive the minimum amount of gossip a population needs to reach consensus and stabilize cooperation. These results concern evolving populations, in which strategies are selected across generations. We study the corresponding question for reinforcement learning agents.

\textbf{Evolutionary Biology.} Reputation has long been studied in evolutionary biology as a mechanism for indirect reciprocity~\cite{nowak_98,nowak_06,milinski_02}, with subsequent work characterizing the assessment rules that sustain cooperation~\cite{ohtsuki_06} and the design of reputation systems~\cite{dellarocas_03}. Much of this theory assumes that reputations are publicly shared. Relaxing this assumption introduces private assessments and, consequently, disagreement. Such disagreement can undermine cooperation~\cite{Hilbe_18}, and no cooperative norm is evolutionarily stable when individuals' opinions are statistically independent~\cite{murase_24}. Gossip provides a mechanism to overcome this informational fragmentation by allowing individuals to learn about others beyond their own direct observations~\cite{sommerfeld_07,feinberg_14}. Kawakatsu et al.~\cite{Kawakatsu_24}, for example, characterize how much gossip is required for a population to reach sufficient consensus to stabilize cooperation. These results are developed primarily for evolutionary populations, where strategies spread through selection across generations. We study a related question in reinforcement learning agents by examining how the diffusion of private assessments through gossip shapes cooperation as agents learn over dynamic communication networks.

%\textbf{Opinion dynamics and evolving networks.} We model gossip with the DeGroot model~\cite{DeGroot_74}, in which each agent repeatedly replaces its opinion with an average of its neighbors' opinions. Its behavior is well understood: repeated averaging aggregates the agents' information and leads to consensus over time~\cite{DeMarzo_03,golub_10}. Evolving networks have also been found to matter for cooperation directly. Outcomes differ qualitatively when ties are rewired on a timescal~\cite{Pacheco_06,Su_23}, and human groups sustain cooperation more readily when they can update their connections~\cite{Rand_11}. To the best of our knowledge, these two strands have not been combined for learning agents.

\textbf{Opinion Dynamics and Evolving Networks.} We model gossip using the DeGroot model~\cite{DeGroot_74}, in which agents repeatedly update their opinions by averaging those of their neighbors. Repeated averaging provides a simple mechanism for aggregating distributed information and, under appropriate conditions, leads opinions toward consensus~\cite{DeMarzo_03, golub_10}. Separately, a large body of work has shown that network dynamics can shape the emergence of cooperation. Cooperation can change qualitatively when social ties are rewired over time~\cite{Pacheco_06,Su_23}, and experiments with human groups show that cooperation is more readily sustained when participants can modify their connections~\cite{Rand_11}. To the best of our knowledge, these two strands have not been brought together in MARL to study how opinions propagate over evolving communication networks and, in turn, shape cooperation among learning agents.
\section{Preliminaries}
\subsection{Social Dilemmas}
\label{sec::social-dilemmas}
We examine social dilemmas using repeated two-player symmetric games, in which each player chooses either to cooperate ($C$) or to defect ($D$). The game matrix is defined as:
\begin{equation}
\label{eq:matrix}
M = \;
\begin{array}{c|cc}
    & C & D \\ \hline
  C & R,\, R & S,\, T \\
  D & T,\, S & P,\, P
\end{array}
\end{equation}
where each entry lists the row player's payoff, followed by the column player's. Mutual cooperation yields the reward $R$ and mutual defection yields the punishment $P$. When one player cooperates and the other defects, the cooperator receives the sucker's payoff $S$ and the defector receives the temptation payoff $T$. In this paper, we focus on the Prisoner's Dilemma, which satisfies $T > R > P > S$, in which defection is a strictly dominant strategy, and mutual defection is the unique Nash equilibrium even though both players would prefer mutual cooperation ($R > P$). Following this convention, we define our Prisoner's Dilemma game matrix as
\begin{equation}
\label{eq:pd}
PD = \;
\begin{array}{c|cc}
    & C & D \\ \hline
  C & 2,\, 2 & 0,\, 3 \\
  D & 3,\, 0 & 1,\, 1
\end{array}
\end{equation}

We extend the two-player PD to $N$ agents as in \cite{LR2}. At the start of each round, the agents are partitioned uniformly at random into neighborhoods of size $k$. When $N$ is not divisible by $k$, the remaining agents form a smaller neighborhood, and a single remaining agent sits out the round. Each agent $i$ chooses one action $a_i \in \{C, D\}$ and plays it against every other agent in its neighborhood $G_i$, receiving the mean of the resulting pairwise payoffs:
\begin{equation}
\label{eq:pd_reward}
r_i^{\mathrm{env}} = \frac{1}{|G_i| - 1} \sum_{j \in G_i \setminus \{i\}} PD_{\mathrm{row}}(a_i, a_j),
\end{equation}
where $PD_{\mathrm{row}}(a_i, a_j)$ is the row player's payoff in Equation~\eqref{eq:pd}. Because defection dominates in every pairwise term, it also dominates at the neighborhood level, so the social dilemma is preserved.

\subsection{Sequential Social Dilemmas}
%Sequential social dilemmas extend social dilemmas from matrix games to Markov environments, where cooperation and defection are expressed through temporally extended behaviors rather than individual actions. We focus on Harvest, which is a partially observable gridworld where $N$ agents move through a 2D map populated with apples. An agent collects an apple by stepping onto its tile and receiving a reward of $1$.  Apples regrow at a rate determined by the local density of remaining apples, which means that if apples are depleted within a region, no more apples will grow until the next episode. Apples also do not spawn on tiles currently occupied by an agent. Therefore, a strategy that aggressively harvests apples drives apple fields to extinction, leading to the Tragedy of the Commons, whereas sustainable harvesting yields higher long-run returns. Each agent is also equipped with a time-out beam that temporarily removes the targeted agent from play for $s$ steps. Their observations are also constrained by a fixed view radius $r$.

Sequential social dilemmas extend social dilemmas from matrix games to Markov environments, where cooperation and defection are expressed through temporally extended behaviors rather than individual actions. We focus on Harvest, a partially observable gridworld in which $N$ agents move through a 2D environment populated with apples. An agent collects an apple by stepping onto its tile, receiving a reward of $1$. Apples regrow at a rate determined by the local density of remaining apples, but cease to regrow in regions where they are completely depleted. They also cannot regrow on tiles occupied by agents. These dynamics create a social dilemma: aggressive harvesting provides immediate individual rewards but can deplete apple fields, producing a Tragedy of the Commons, whereas sustainable harvesting preserves the shared resource and yields higher long-term returns. Agents can also temporarily remove one another from play for $s$ steps using a time-out beam, while their observations are restricted to a fixed view radius $r$.

%Prior work on Harvest reports a three-stage dynamic~\cite{Leibo_17}. Early in training, agents enter an exploration phase. Then, agents overharvest the apple field, leading to the Tragedy of the Commons phase, in which collective returns collapse. In the final phase, agents learn to use the time-out beam strategically so that the smaller surviving group can sustainably exploit the regrowing resource. Sustainability in Harvest thus emerges not from voluntary restraint but from population control through violence, an outcome that motivates our interest in reputation as an alternative coordination mechanism.

Prior work on Harvest identifies a three-stage dynamic~\cite{Leibo_17}. Agents initially enter an exploration phase, followed by a Tragedy of the Commons phase in which widespread overharvesting depletes the field and causes collective returns to collapse. At the end, agents learn to use the time-out beam strategically, reducing the number of active agents and allowing the remaining population to harvest the regrowing resource. Sustainability in Harvest therefore emerges not through voluntary restraint, but through the exclusion of competing agents. This dynamic motivates reputation as an alternative mechanism for inducing sustainable behavior.
\section{Methods}
\label{sec:method}

\begin{table*}[t]
\centering
\begin{tabular}{clc}
\hline
Symbol & Description & Range/Type \\
\hline
$N$              & Number of agents                          & $\mathbb{Z}^+$ \\
$\sigma_{ij}^t$ & Observation signal of agent $i$ towards agent $j$ at time $t$ & $\{-1, 0, 1\}$ \\
$\omega_{ij}^t$ & Raw opinion of agent $i$ towards agent $j$ at time $t$ & $[-1, 1]$\\
$\alpha$ & Exponential moving average factor & $[0, 1]$\\
$\mu_j^t$ & Reputation entry of agent $j$ at time $t$ & $[-1, 1]$ \\
$E_j^t$ & Reputation over a sliding window of agent $j$ at time $t$ & $[-1, 1]$\\
$K$ & Sliding window size for calculating $E_j^t$ & $\mathbb{Z}^+$\\
$\rho$ & Broadcast radius & $\mathbb{R}_{\geq 0} \cup \{\infty\}$\\
$o_{ij}^t$ & Opinion of agent $i$ towards agent $j$ at time $t$ after broadcast & $[-1, 1]$\\
$r$ & Observation radius & $\mathbb{Z}^+$\\
$s_i^t$          & Local observation of agent $i$ at time $t$ & $\{0,\dots,255\}^{(2r + 1) \times (2r + 1) \times 3}$\\
$R^{(i, t)}$            & Reward of agent $i$ at time $t$             & $\mathbb{R}$ \\
$r_{ex}^{(i, t)}$            & External reward of agent $i$ at time $t$& $\mathbb{R}^+$ \\
$\gamma$         & Discount factor                            & $[0, 1)$ \\
$\beta$ & Strength of reputation & $\mathbb{R}^+$\\
\hline
\end{tabular}
\caption{Description of Variables}
\label{tab:variables}
\end{table*}
%We are interested in emergent behavior in standard reinforcement learning agents, and in particular in whether such agents can learn to act on a reputation that is not given to them but communicated through gossip. We therefore do not hand-engineer prosocial incentives into the agents. Instead, agents judge one another using a simple, fixed rule based on what they directly observe, pass these opinions on to their neighbors through gossip, and are rewarded by the resulting reputation. Therefore, any prosocial strategy must be discovered by the RL algorithm from this reward signal. To this end, we propose a flexible, observation-based reputation framework in which the reach of gossip is a tunable parameter, spanning a spectrum of real-world reputation systems from purely first-hand opinion to population-wide broadcast. We describe its design in the following subsection.

We are interested in the behavior that emerges from standard reinforcement learning agents, and particularly in whether they can learn to cooperate when reputational information is acquired through direct observation and gossip. We do not hand-engineer prosocial strategies into the agents. Instead, agents form opinions of others using simple, fixed assessment rules based on their direct observations, exchange these opinions with their neighbors through gossip, and learn from the reputational information that results. An agent's own reputation shapes its reward, while its opinions of others can inform its decisions. Prosocial strategies must therefore be discovered by the RL algorithm in response to these reputation signals. To study how the diffusion of reputational information affects this learning process, we introduce a flexible, observation-based reputation framework in which the reach of gossip is controlled by a tunable parameter, spanning no information sharing, local gossip, and population-wide broadcast. We describe the framework in the following subsection.

\begin{algorithm}[t]
\caption{Reputation Framework}
\label{alg:reputation}
\DontPrintSemicolon
\SetAlgoLined
 
\KwData{Agent set $\mathcal{A}=\{1,\dots,N\}$, horizon $T$, EMA factor $\alpha$,
        window $K$, broadcast radius $\rho$, view radius $r$,
        reputation strength $\beta$, distance $d(\cdot,\cdot)$}
\KwResult{Reputation $E$}
 
$o_{ij}^{0} \leftarrow 0$ \quad $\forall\, i \neq j$; \; $\mu_{j}^{0} \leftarrow 0$ ; \quad $\mathcal{H}_j \leftarrow \textsc{Deque}(K)$; \quad
$E_j^0 \leftarrow 0$\quad $\forall\ j \in \mathcal{A}$\;

\For{$t \leftarrow 1$ \KwTo $T$}{
  \tcp{Opinion formation}
  \ForEach{$i \in \mathcal{A}$}{
    $s_i^t \leftarrow \mathrm{observe}(i, r)$ \nllabel{ln:observe}
    \ForEach{$j \in \mathcal{A} \setminus \{i\}$}{
      \lIf{$j$ prosocial in $s_i^t$}{$\sigma_{ij}^t \leftarrow +1$\nllabel{ln:sigpro}}
      \lElseIf{$j$ antisocial in $s_i^t$}{$\sigma_{ij}^t \leftarrow -1$}
      \lElse{$\sigma_{ij}^t \leftarrow 0$\nllabel{ln:signull}}
      \lIf{$\sigma_{ij}^t \neq 0$}{
        $\omega_{ij}^t \leftarrow \alpha o_{ij}^{t-1} + (1-\alpha)\sigma_{ij}^t$\nllabel{ln:ema}}
      \lElse{$\omega_{ij}^t \leftarrow o_{ij}^{t-1}$\nllabel{ln:noema}}
    }
  }
  \tcp{Broadcast}
  \ForEach{$i \in \mathcal{A},\ j \in \mathcal{A}\setminus\{i\}$}{
    $\mathcal{N}_{i \to j}^{\rho}(t) \leftarrow \{x \in \mathcal{A}\setminus\{j\} : d(i,x) \leq \rho\}$\nllabel{ln:nbhd}\;
    $o_{ij}^t \leftarrow \frac{1}{|\mathcal{N}_{i \to j}^{\rho}(t)|}
      \sum_{x \in \mathcal{N}_{i \to j}^{\rho}(t)} \omega_{xj}^{t}$\nllabel{ln:pool}\;
  }
  \tcp{Temporal aggregation}
  \ForEach{$j \in \mathcal{A}$}{
    \eIf{$\exists\, i \neq j:\ \sigma_{ij}^t \neq 0$\nllabel{ln:cond}}{
      $\mu_j^t \leftarrow \frac{1}{N-1}\sum_{i \neq j} \omega_{ij}^t$\nllabel{ln:entry}\;
      $\mathcal{H}_j \leftarrow \mathcal{H}_j \cup \mu_j^t$\nllabel{ln:append}\;
      $L \leftarrow min(|\mathcal{H}_j|, K)$ \;
      $E_j^t \leftarrow \frac{1}{L}\sum_{k = 0}^{L - 1}\mathcal{H}_j[k]$\nllabel{ln:window}\;
    }{
      $E_j^t \leftarrow E_j^{t-1}$\nllabel{ln:carry}\;
    } 
  }
 % \tcp{Reputation-shaped reward}
  %\lForEach{$i \in \mathcal{N}$}{$R^{(i,t)} \leftarrow r_{ex}^{(i,t)} + \beta E_i^t$\nllabel{ln:reward}}
}

%\Return $R^t$\;
\Return $E$\;
% E^t,\ o^t,\ R^t$\;
\end{algorithm}

% We are also interested in investigating emergent behavior in standard reinforcement learning agents, particularly whether such agents can learn to use reputation. Therefore, we do not hand-engineer prosocial incentives into the agents; instead, we rely on the RL algorithm to discover prosocial strategies through reward signals shaped by reputation.
\begin{figure*}[htbp]
    \centering
   \includegraphics[width=0.9\linewidth]{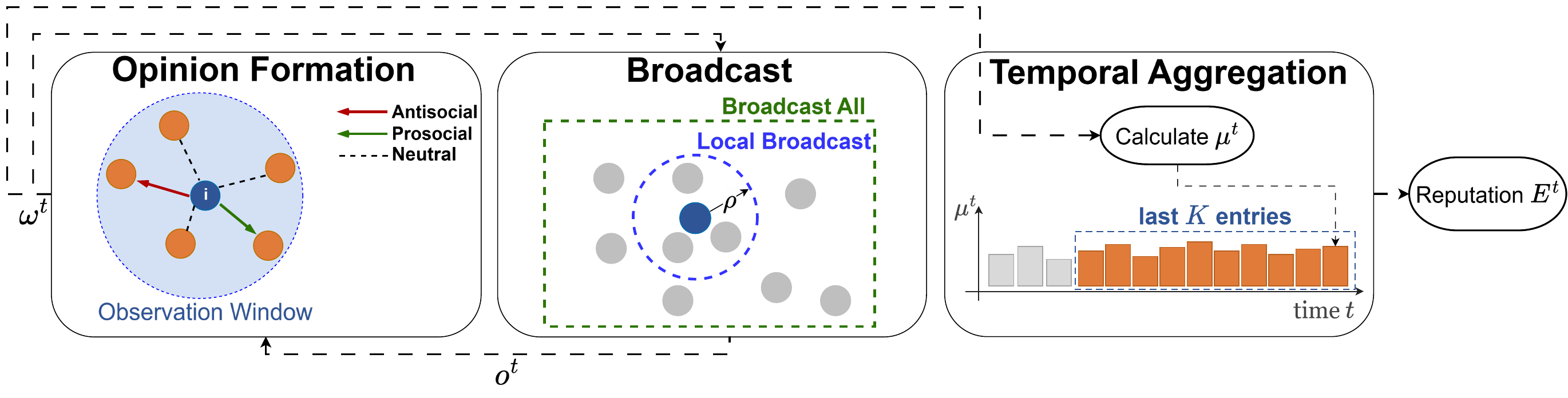}
         % \includesvg[width=0.9\linewidth]{Figures/main_figure.svg}

    \caption{The reputation architecture includes an \textit{opinion formation} phase through agents' local observations, a \textit{temporal aggregation} phase that calculates the current step reputation, and a \textit{broadcast} phase that spreads reputation information.
    }
   \label{fig::main}
\end{figure*}
\subsection{Reputation Framework}
%Our framework is based on three core mechanisms: local observation, parameterized broadcasting, and temporal aggregation. First, local observation enforces decentralized opinion formation, as each individual's judgments are constrained by their view lengths. Second, we model a range of reputation systems by parameterizing the broadcast radius, which encompasses protocols that correspond to real-world scenarios: centralized rating systems and word-of-mouth. Third, opinions after broadcast are aggregated and enable agents to maintain persistent memories of past interactions while still allowing reputation to recover over time. 

Our framework consists of three core mechanisms: local observation, parameterized broadcasting, and temporal aggregation. First, agents form private opinions from local observations, allowing different agents to hold different assessments of the same individual. Second, agents exchange these opinions through a parameterized broadcasting mechanism, where the broadcast radius controls how far reputational information propagates. Varying this radius spans settings ranging from private, first-hand information to local word-of-mouth and population-wide broadcasting. Third, the private opinions from the first step are aggregated over time into reputation signals, providing a persistent memory of past behavior while allowing reputations to adapt as agents' behavior changes. Algorithm \ref{alg:reputation} provides the pseudocode for the framework, which we describe in detail below.

\textbf{Opinion Formation.} Let $\mathcal{N}=\{1,\dots,N\}$ denote the set of agents. At each time step $t$, every agent $i\in\mathcal{N}$ maintains an opinion $o_{ij}^t$ of every other agent $j\neq i$. These opinions are updated from direct observations within $i$'s view window (line~\ref{ln:observe}). When $i$ observes $j$ engaging in behavior that is identified as prosocial or antisocial in the given environment, it generates an observation signal $\sigma_{ij}^t$ that is positive or negative, respectively. If no such behavior is observed, $\sigma_{ij}^t=0$ (lines~\ref{ln:sigpro}--\ref{ln:signull}). Agent $i$ incorporates a nonzero signal into its raw opinion $\omega_{ij}^t$ using an exponential moving average (line~\ref{ln:ema}), where $\alpha\in[0,1]$ controls the weight assigned to its previous opinion relative to the new observation. When $\sigma_{ij}^t=0$, the previous opinion is carried forward unchanged (line~\ref{ln:noema}), ensuring that the absence of new evidence does not alter the agent's assessment. The behaviors considered prosocial or antisocial are defined by the context of each environment. Because agents assess only behavior they directly observe, events outside agent $i$'s view do not affect $\sigma_{ij}^t$, ensuring that opinions are formed from strictly local information.

\textbf{Broadcast.} The reputation systems we consider differ in how raw opinions are shared among agents. After incorporating its private observation signal, each agent exchanges its raw opinions with its communication neighbors through a DeGroot-style averaging step \citep{DeGroot_74}. The communication neighborhood of agent $i$ consists of all agents within distance $\rho$ of $i$ (line~\ref{ln:nbhd}), where $d(\cdot,\cdot)$ denotes a distance function. Because $i$ belongs to its own neighborhood for any $\rho \geq 0$, its own assessment is always included. The post-broadcast opinion $o_{ij}^t$ is the average opinion about agent $j$ among agents in $i$'s communication neighborhood (line~\ref{ln:pool}). Thus, $\rho$ controls how far reputational information propagates, while DeGroot averaging determines how agents combine their neighbors' opinions.

%We consider three broadcasting protocols as we vary $\rho$. \textbf{Broadcast-to-all} ($\rho = \infty$) overwrites the opinions of all agents, modeling centralized rating systems such as those used in online marketplaces, where every buyer observes the same rating of a seller; new buyers' opinions are likely formed by the given rating, and buyers rate sellers after each purchase to shape the seller's future reputation. \textbf{Broadcast-to-none} ($\rho = 0$) leaves all opinions unchanged, modeling a fully private information setting in which no aggregation channel exists. \textbf{Local broadcast} ($0 < \rho < \infty$) updates only opinions in a spatial neighborhood of $j$, modeling word-of-mouth reputation in which information about an individual reaches their immediate social neighborhood (e.g., coworkers gossiping about a manager) but not the periphery. By sweeping over $\rho$, we could model varying network sizes while holding opinion formation and aggregation fixed.

We consider three broadcasting protocols by varying $\rho$. Broadcast All ($\rho=\infty$) shares opinions across the entire population, approximating centralized rating systems such as those used in online marketplaces, where reputational information is globally accessible. \textbf{No broadcast} ($\rho=0$) leaves each agent's opinions unchanged, corresponding to a fully private setting in which agents rely only on their own observations. \textbf{Local broadcast} ($0<\rho<\infty$) shares opinions only within local communication neighborhoods, modeling word-of-mouth reputation in which information propagates among nearby or socially connected individuals without necessarily reaching the entire population. By varying $\rho$, we control the reach of reputational information while holding opinion formation and aggregation fixed.

%\textbf{Reputation Through Temporal Aggregation.} Whenever at least one opinion about $j$ has been updated at step $t$ (line~\ref{ln:cond}), a new reputation \emph{entry} $\mu_j^t$ is computed as the population mean of opinions about $j$ (line~\ref{ln:entry}) and appended to $j$'s entry list $H_j$ (line~\ref{ln:append}). The reputation $E_j^t$ is the average of the $K$ most recent entries of $H_j$ (line~\ref{ln:window}). Crucially, an entry is appended \emph{only} when some agent has actually observed $j$; on steps where no one does, nothing enters the window and the previous reputation is carried forward (line~\ref{ln:carry}). The window, therefore, advances at the rate at which behavior is observed rather than with each timestep. The window smooths short-term fluctuations while still allowing reputation to track behavioral change. 

\textbf{Reputation Through Temporal Aggregation.} Whenever at least one opinion about agent $j$ is updated at step $t$ (line~\ref{ln:cond}), a new reputation entry $\mu_j^t$ is computed by averaging the population's opinions of $j$ (line~\ref{ln:entry}) and appended to its history $H_j$ (line~\ref{ln:append}). Agent $j$'s reputation $E_j^t$ is then defined as the average of the $K$ most recent entries in $H_j$ (line~\ref{ln:window}). If $j$ is not observed at step $t$, no new entry is added, and its previous reputation is carried forward (line~\ref{ln:carry}). Thus, the aggregation window advances only when new behavioral evidence is observed, smoothing short-term fluctuations while allowing reputation to adapt to changes in behavior.

% Each agent maintains its true reputation $E_j^t$ and receives it as an intrinsic reward scaled by $\beta$, added to the external game-playing reward (line~\ref{ln:reward}), while other agents can access this information only through broadcasts.

\section{Experiments}
\label{sec:experiments}

We evaluate our reputation framework under the three broadcasting protocols introduced in Section~\ref{sec:method}: \textbf{No Broadcast}, \textbf{Local Broadcast}, and \textbf{Broadcast All}. We compare these methods against a \textbf{No Reputation} baseline, in which agents are trained using only the extrinsic environment reward and receive no reputational information. This baseline corresponds to standard independent deep Q-learning, allowing us to isolate the effect of introducing reputation and its diffusion through gossip.

\subsection{Environment}
\label{sec:env}
%\textbf{Prisoner's Dilemma.} We use the $n$-player Prisoner's Dilemma with random neighborhoods described in Section \ref{sec::social-dilemmas} with $N = 15$ agents and neighborhoods of size $k = 5$, so that every round partitions the population into three tables and no agent sits out. Payoffs are $T = 3$, $R = 2$, $P = 1$, and $S = 0$, and each agent receives the mean of its pairwise payoffs against its four neighbors (Equation~\ref{eq:pd_reward}). An episode lasts $1000$ steps, and the tables are uniformly at random redrawn at the start of each round. At each round, agent $i$ observes the opinions of its upcoming neighbors, sorted in ascending order, along with its own reputation, and chooses whether to cooperate or defect.

\subsubsection{Prisoner's Dilemma:} We use the $n$-player Prisoner's Dilemma with random neighborhoods described in Section~\ref{sec::social-dilemmas}. We consider $N=15$ agents and neighborhoods of size $k=5$, partitioning the population into three tables at each round. Payoffs are $T=3$, $R=2$, $P=1$, and $S=0$, and each agent receives the mean of its pairwise payoffs against the other four agents at its table (Equation~\ref{eq:pd_reward}). Each episode lasts $1000$ steps, with tables redrawn uniformly at random before each round. Before choosing whether to cooperate or defect, agent $i$ observes its own reputation and the opinions it holds of its upcoming tablemates, sorted in ascending order.

%\textbf{Harvest.} We recreate the Harvest environment from an open-source GitHub repository~\cite{SSDOpenSource}. We run the experiments with $N = 10$ agents, and each agent's view length has a radius $r = 5$. The game environment is depicted in Figure \ref{fig::game_map}.  Tiles are rendered as follows: empty tiles are black, walls grey, apples green, the firing beam yellow, and the observing agent itself blue. Other agents are colored by the observer's opinion of them: yellow if $o_{ij}^t \geq 0$ and red otherwise. Reputation thus enters the agent's observation as a perceptual feature, and the broadcast radius governs how much reputational information each agent perceives.

\subsubsection{Harvest:} 
\begin{figure}[h!]
    \centering

        \includegraphics[width=0.45\textwidth]{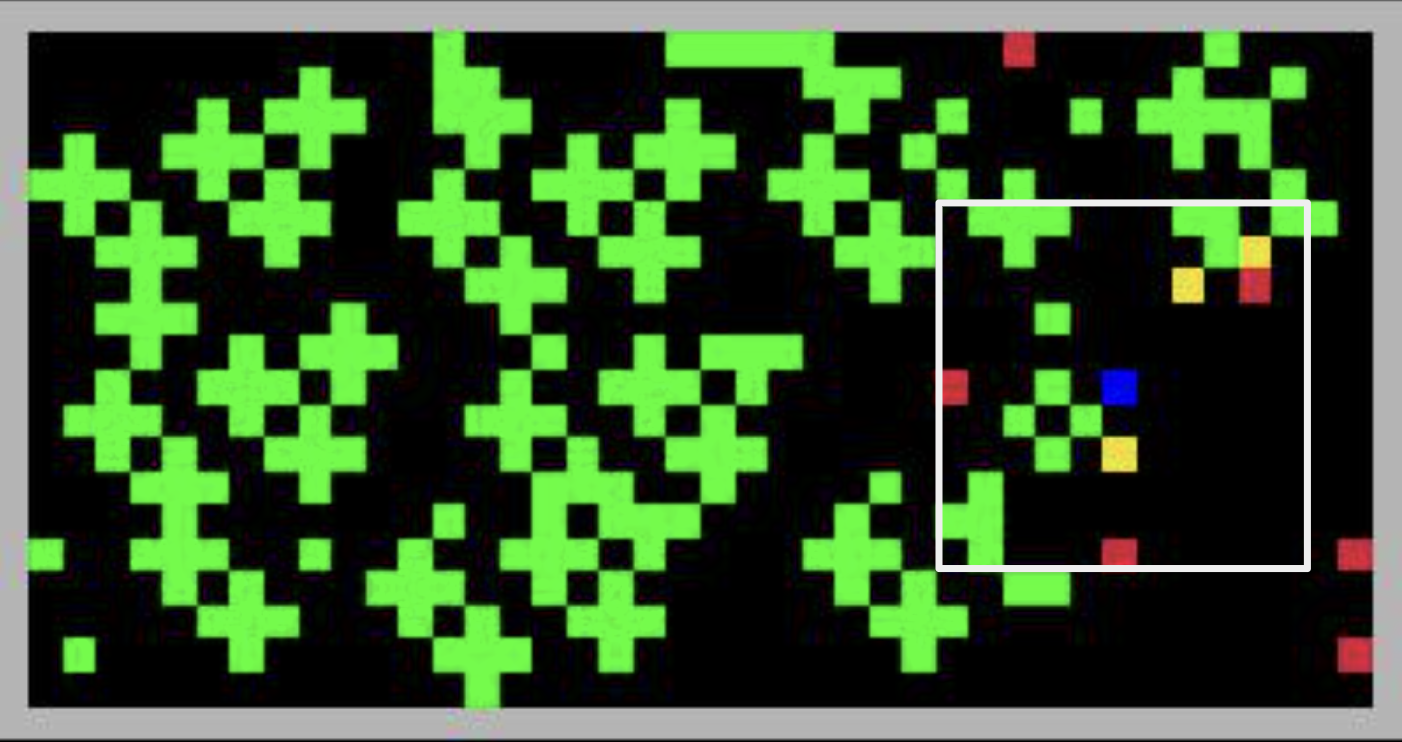}

    \caption{%The Harvest environment with reputation. The full image shows the global map, while the white box shows the image observation received by the observing (blue) agent; the white box itself is not part of the observation.
    The Harvest environment with reputation. The full image depicts the global environment, while the white box indicates the local observation available to the observing (blue) agent. The white border is shown only for visualization and is not part of the agent's observation.
    }
   \label{fig::game_map}
\end{figure}

We recreate the Harvest environment using an open-source implementation~\cite{SSDOpenSource}. We consider $N=10$ agents with an observation radius of $r=5$. Figure~\ref{fig::game_map} depicts the environment. Empty tiles are shown in black, walls in grey, apples in green, and the observing agent in blue. Other agents are colored according to the observer's opinion of them: yellow if $o_{ij}^t \geq 0$ and red otherwise. Opinions enter each agent's observation as perceptual features, while the broadcast radius $\rho$ controls how far they propagate.

%\textbf{Trainning.} We train each agent independently with its own Deep Q-Network~\cite{DQN_15}. To couple the reputation mechanism of Section~\ref{sec:method} to learning, we treat reputation as an intrinsic reward. When agent $i$ takes action $a_i$ at state $s_i$ at step $t$, the reward received is:
\textbf{Training.} We train each agent independently using a Deep Q-Network (DQN)~\cite{DQN_15}. To incorporate reputation into the learning process, we augment the environment reward with an intrinsic reward determined by the agent's reputation. Specifically, when agent $i$ takes action $a_i$ in state $s_i$ at step $t$, its learning reward is:
\begin{equation}
    R^{(i,t)}(s_i, a_i) \;=\; r_{ex}^{(i,t)}(s_i, a_i) \;+\; \beta E_i^t,
\end{equation}
%where $r_e(s_i, a_i)$ is the extrinsic, standard environment reward for either matrix game payoffs or harvesting apples, $R_i^t$ is agent $i$'s current reputation, and $\beta$ controls the strength of the reputation signal. Since $R_i^t$ is an average of opinions and thus bounded in $[-1, 1]$, we set $\beta = 1$ in our experiments. Each agent independently optimizes the discounted sum $\sum_{t} \gamma^t r_i^t$.
where $r_{ex}(s_i,a_i)$ is the extrinsic reward provided by the environment (matrix-game payoff in the Prisoner's Dilemma and apple collection in Harvest); $E_i^t$ is agent $i$'s current reputation; and $\beta$ controls the contribution of reputation to the learning reward. Because $E_i^t$ is an average of opinions and is therefore bounded in $[-1,1]$, we set $\beta=1$ in all experiments. Each agent independently learns a policy that maximizes the discounted sum $\sum_{t} \gamma^t R^{(i,t)}$.

\subsection{Opinion Formation}
We formalize the opinion formation rule introduced in Section~\ref{sec:method} for the two games below.

\subsubsection{Prisoner's Dilemma:}
\begin{enumerate}
    \item If $i$ observes $j$ cooperate, then $\sigma_{ij}^t = +1$
    \item If $i$ cooperates and observes $j$ defect, then $\sigma_{ij}^t = -1$
\end{enumerate}
Because tablemates play against one another, an observer's assessment reflects how a neighbor's action affected the observer. The rule therefore condemns exploitation rather than defection. A player who defects against a cooperating observer exploits that observer and is assessed negatively, whereas mutual defection results in no assessment. Cooperation is always assessed positively because it does not exploit the observer. Consequently, the same action can produce different assessments across tablemates: a player who defects against the entire table is assessed negatively by those who cooperated but receives no assessment from those who also defected. Opinions therefore depend on agents' local experiences.%, allowing different observers of the same action to reach different assessments.
%Because neighbors play against one another, an observer's assessment of a neighbor reflects how that tablemate treated the observer. Therefore, the rule condemns exploitation rather than defection. A player who defected against a cooperating observer exploited it and is condemned, whereas a player who defected against an observer that also defected exploited no one in that exchange, and the observer abstains. Cooperation harms no one and is always credited. This rule also makes assessment depend on the observer's own experience: a single action, played against the whole table, is exploitation to the tablemates who cooperated and not to those who defected, so observers of the same act can reach different assessments. This introduces partial information, where the agent's assessment depends on its own local circumstances.

\subsubsection{Harvest:}
\begin{enumerate}
    \item If $i$ observes $j$ consume one of the last two apples within $i$'s view, then $\sigma_{ij}^t = -1$. If $i$ observes $j$ adjacent to such an apple but choosing not to consume it, then $\sigma_{ij}^t = +1$.
    \item If $i$ observes no apple around, then for all neighbors $j$ within observation $\sigma_{ij}^t = -1$.
    \item If $i$ observes $j$ firing the tagging beam at any agent $k$ with reputation $E_k^t \geq 0$, then $\sigma_{ij}^t = -1$.
\end{enumerate}
%The first and second rules treat depletion of the shared resource as antisocial and restraint as prosocial. The third rule treats aggression toward agents in good standing as antisocial, mirroring real-world norms in which unprovoked violence against an innocent party is widely viewed as unforgivable~\cite{cushman_08}. Because no central authority enforces norms in this environment, tagging a low-reputation agent yields neither a reputational reward nor a penalty, which is a design that mirrors altruistic punishment~\cite{fehr_02, fehr_04}. Under this design, an agent bears a personal cost to sanction defectors, giving up the opportunity to collect apples during the steps spent firing a time-out beam, without external compensation.

The first rule treats exploitation of the shared resource as antisocial and restraint as prosocial. The second rule treats resource depletion as a shared, global responsibility. The third treats aggression (i.e., beaming) toward agents in good standing as antisocial, reflecting social norms against unprovoked harm~\cite{cushman_08}. In contrast, tagging a low-reputation agent carries neither a reputational reward nor a penalty, allowing agents to sanction antisocial behavior without external compensation. This design resembles altruistic punishment~\cite{fehr_02,fehr_04}: an agent bears the opportunity cost of sanctioning a defector, forgoing apple collection while firing the time-out beam, without receiving a direct reward for doing so.

\subsection{Evaluation Metrics}
\subsubsection{Prisoner's Dilemma:} We evaluate the average number of cooperation steps in each episode, formally:
\begin{equation}
    C = \frac{1}{N} \sum_{i=1}^N \sum_{t=1}^T \mathbb{I}\{\text{agent } i \text{ cooperates on step } t\} 
\end{equation}

\subsubsection{Harvest:} We adopt the evaluation metrics from P\'erolat et al.~\cite{Leibo_17} and additionally evaluate punishment accuracy. Let $A_i \coloneq \sum_{t=1}^T r_{ex}^{(i,t)}$ denote the total extrinsic reward received by agent $i$ over an episode of horizon $T$, and let $N$ denote the number of agents. We consider five metrics: \textit{Utility} ($U$) is the average return per agent; \textit{Sustainability} ($S$) is the average time at which extrinsic rewards are collected; \textit{Equality} ($E$) is defined as one minus the Gini inequality index~\cite{Sitthiyot_20}; \textit{Peacefulness} ($P$) is the average number of agents that remain untagged at each step; and \textit{Punishment Accuracy} ($PA$) is the fraction of beam uses that target an agent with negative reputation, measuring how consistently punishment is directed toward low-reputation agents. Formally,

%We adopt the evaluation metrics of P\'erolat et al.~\cite{Leibo_17}. Let $A_i \coloneq \sum_{t=0}^T r_{ex}^{(i, t)}$ be the total extrinsic reward received by agent $i$ over one episode, where $T$ is the horizon. Let $N$ be the total number of agents. We define four metrics: \textbf{Utility} ($U$) measuring the average return per agent, \textbf{Sustainability ($S$)} measuring the average time at which extrinsic rewards are collected, \textbf{Equality ($E$)} given by one minus the Gini inequality index\cite{Sitthiyot_20}, \textbf{Peacefulness $(P)$} measuring the average number of agents that are untagged at any given step, and \textbf{Punishment accuracy}($PA$) measuring the fraction of beam uses that correctly tag an agent whose reputation is below zero, i.e., how reliably agents direct punishment at low-reputation agents. Formally:
\begin{align*}
    U &\;=\; \frac{1}{N} \sum_{i=1}^N A_i, \\[4pt]
    S &\;=\; \frac{1}{N} \sum_{i=1}^N t_i, \quad \text{where } t_i \coloneq \mathbb{E}\!\left[t \,\big|\, r_{ex}^{(i,t)} > 0\right], \\[4pt]
    E &\;=\; 1 - \frac{\sum_{i=1}^N \sum_{j=1}^N \left| A_i - A_j \right|}{2 N \sum_{i=1}^N A_i}, \\[4pt]
    P &\;=\; \frac{NT - \sum_{i=1}^N \sum_{t=1}^T \mathbb{I}\{\text{agent } i \text{ timed-out on step } t\}}{T} \\[4pt]
        PA &\;=\; \frac{\sum_{t=1}^{T} \sum_{i=1}^{N} \sum_{j \neq i} \mathbb{I}\{\text{agent } i \text{ tags agent } j \text{ on step } t\}\; \mathbb{I}\{E_j^{t-1} < 0\}}{\sum_{t=1}^{T} \sum_{i=1}^{N} \sum_{j \neq i} \mathbb{I}\{\text{agent } i \text{ tags agent } j \text{ on step } t\}}
\end{align*}
\begin{figure}[h!]
    \centering

    \includegraphics[width=0.45\textwidth]{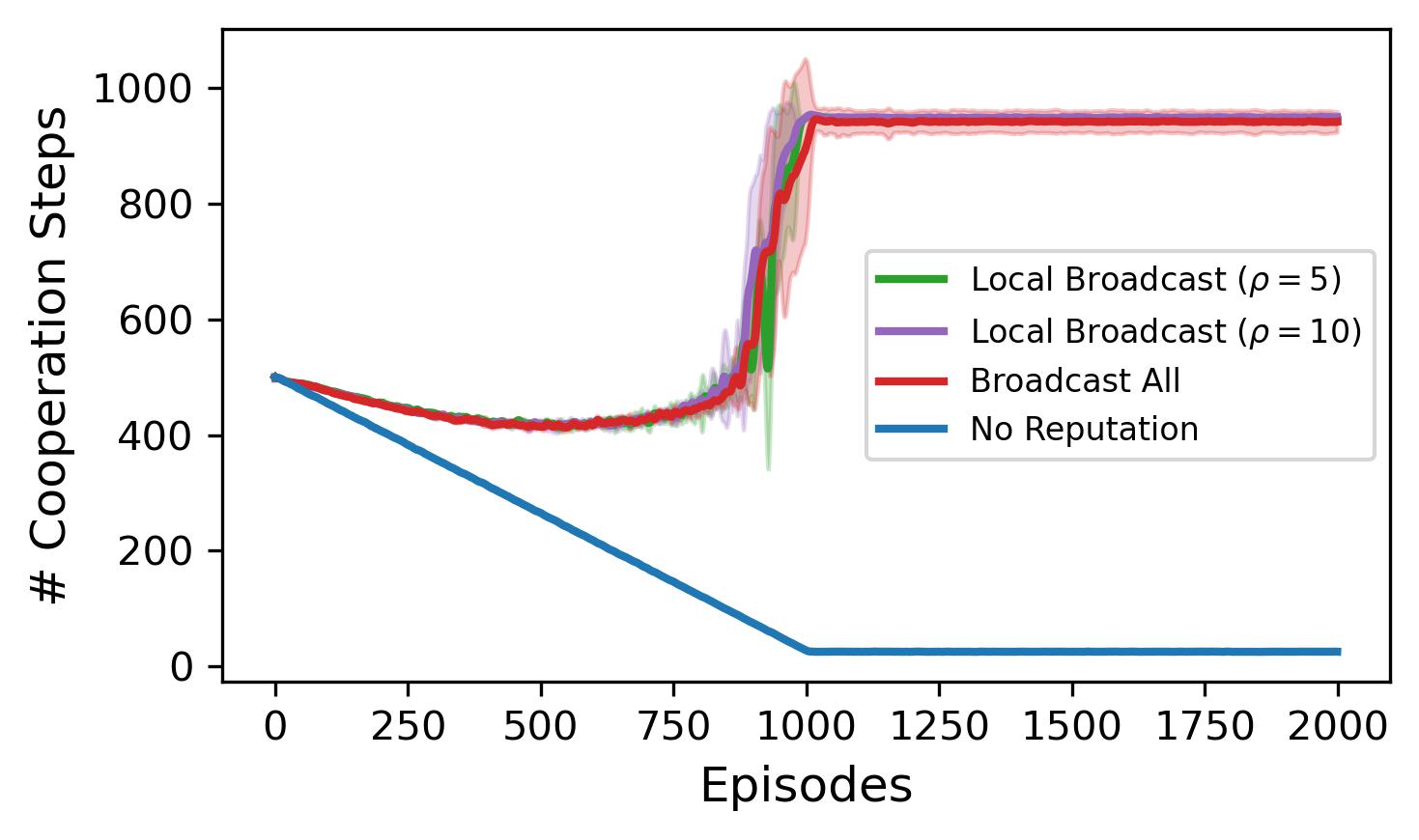}

    \caption{%Average return per agent for three reputation methods and the baseline throughout the trainning. All reputation methods outperforms the baseline, with the broadcast to all method performing the best and local broadcast achieving intermediate results between broadcast to all and no broadcast.
    Average number of cooperation steps per agent throughout training for the three reputation methods and No Reputation in the Prisoner's Dilemma. Reputation methods outperform No Reputation.
    }
    \label{fig::pd}
\end{figure}
\begin{figure}[h!]
    \centering

    \includegraphics[width=0.45\textwidth]{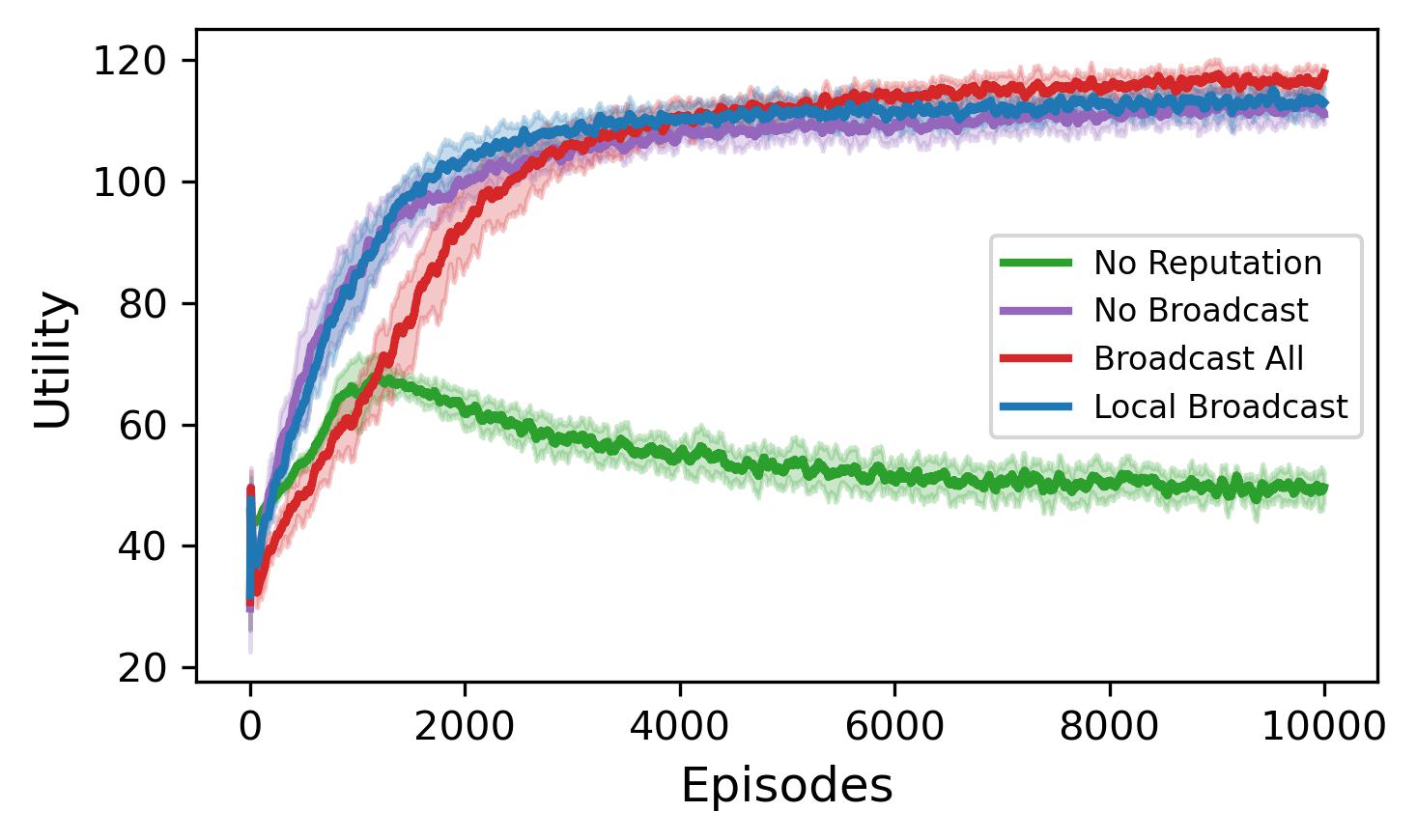}

    \caption{%Average return per agent for three reputation methods and the baseline throughout the trainning. All reputation methods outperforms the baseline, with the broadcast to all method performing the best and local broadcast achieving intermediate results between broadcast to all and no broadcast.
    Average return per agent throughout training for the three reputation methods and No Reputation in Harvest. The results show that reputation methods outperform No Reputation, with Broadcast All achieving the highest return, followed by Local Broadcast and No Broadcast.
    }
    \label{fig::harvest_u}
\end{figure}

\begin{figure}[h!]
    \centering
    
    \includegraphics[width=0.45\textwidth]{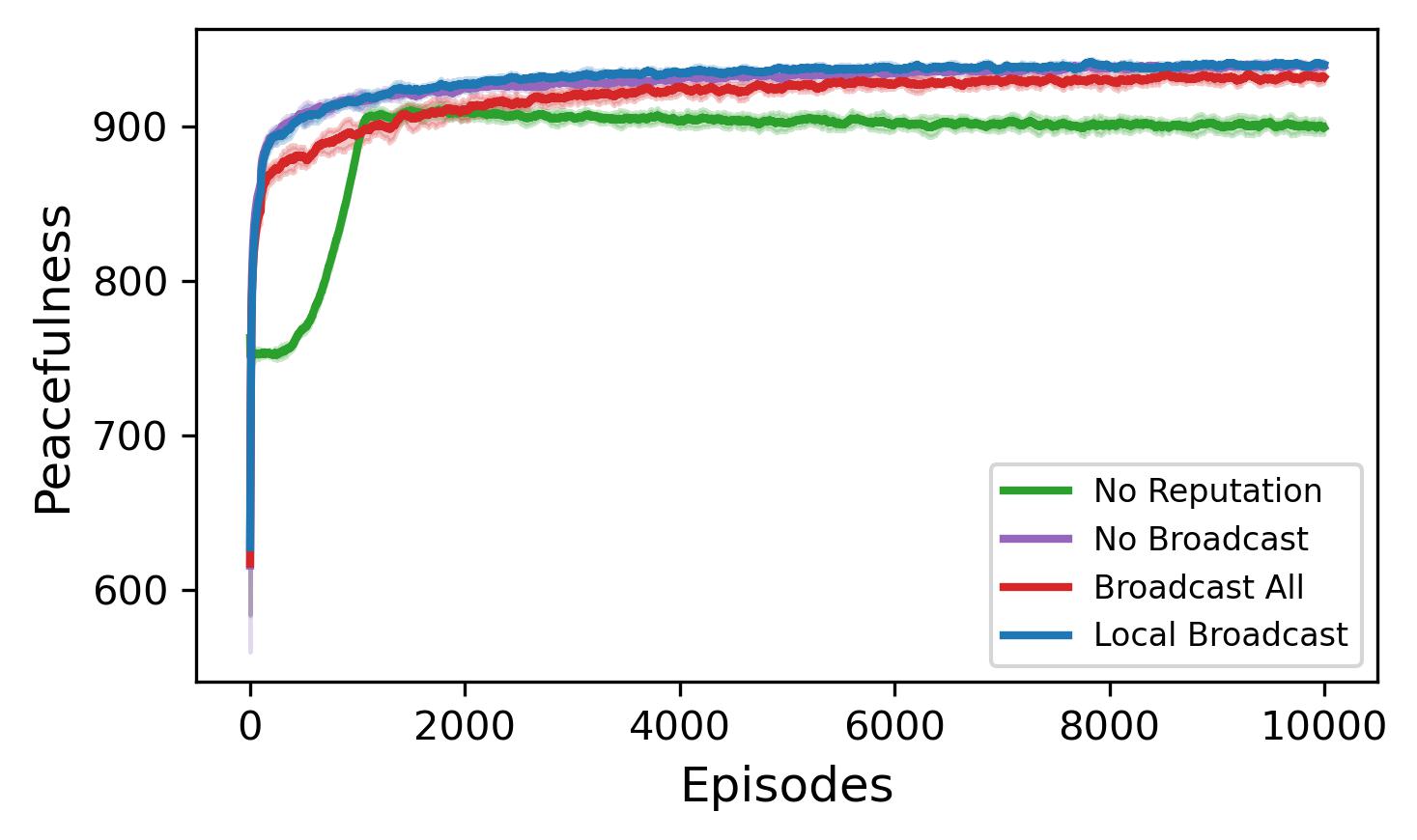}
    
    \caption{%Peacefulness metric for three reputation methods and baseline throughout trainning. Broadcast to all method has slightly worse peacefulness than the other two reputation methods.
    Peacefulness throughout training for the three reputation methods and the No Reputation baseline in Harvest. All reputation methods are more peaceful than the baseline, although Broadcast All achieves slightly lower peacefulness than No Broadcast and Local Broadcast.
    }
    \label{fig::harvest_p}
\end{figure}

\begin{figure}[h!]
    \centering
    
    \includegraphics[width=0.45\textwidth]{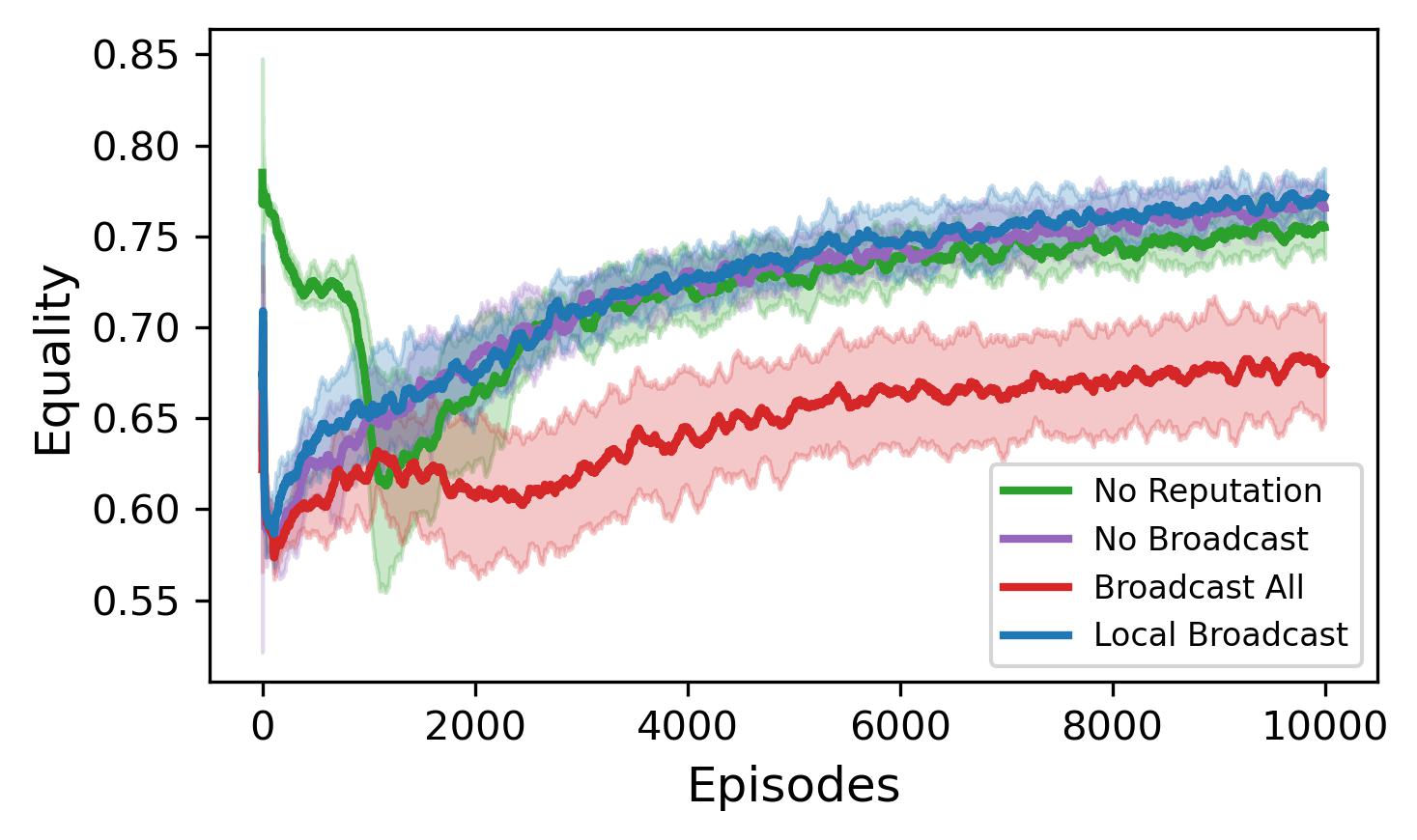}

    \caption{%Equality metric for three reputation methods and baseline throughout trainning. Broadcast to all method sacrifices equality in exchange of higher sustainability and utility.
    Equality throughout training for the three reputation methods and No Reputation in Harvest. Broadcast All exhibits a trade-off, achieving higher sustainability and utility but lower equality compared with the baseline.
    }
    \label{fig::harvest_e}
\end{figure}

\begin{figure}[h!]
    \centering

    \includegraphics[width=0.45\textwidth]{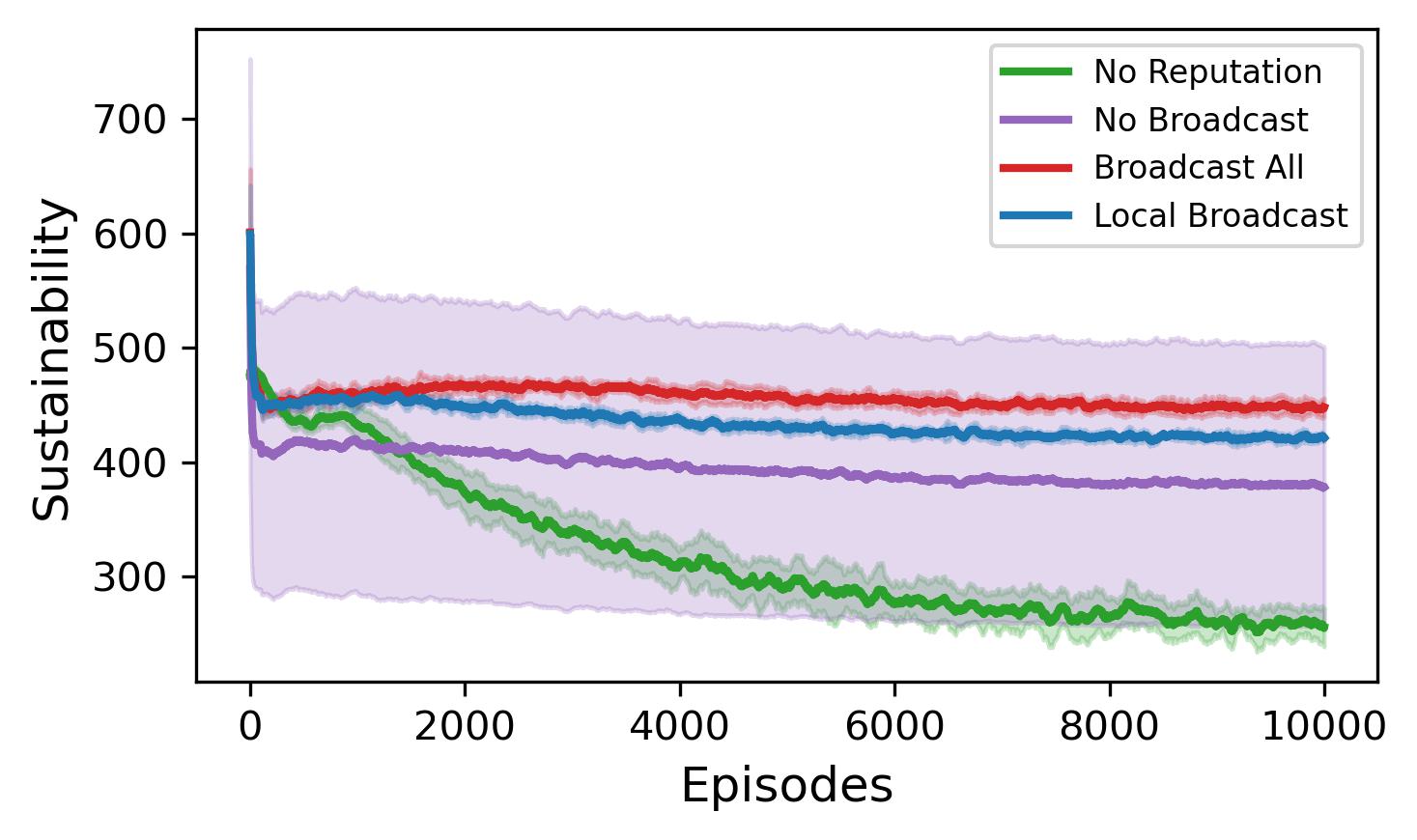}

    \caption{%Sustainability metric for three reputation methods and baseline throughout trainning. Metric increases in proportion to the broadcast radius, and the no broadcast method exhibits very high variation across runs.
    Sustainability throughout training for the reputation methods and No Reputation in Harvest. Reputation improves sustainability relative to the baseline, with sustainability increasing with the broadcast radius. No Broadcast also exhibits substantially greater variation across runs.
    }
    \label{fig::harvest_s}
\end{figure}

\begin{figure}[h!]
    \centering
    \includegraphics[width=0.45\textwidth]{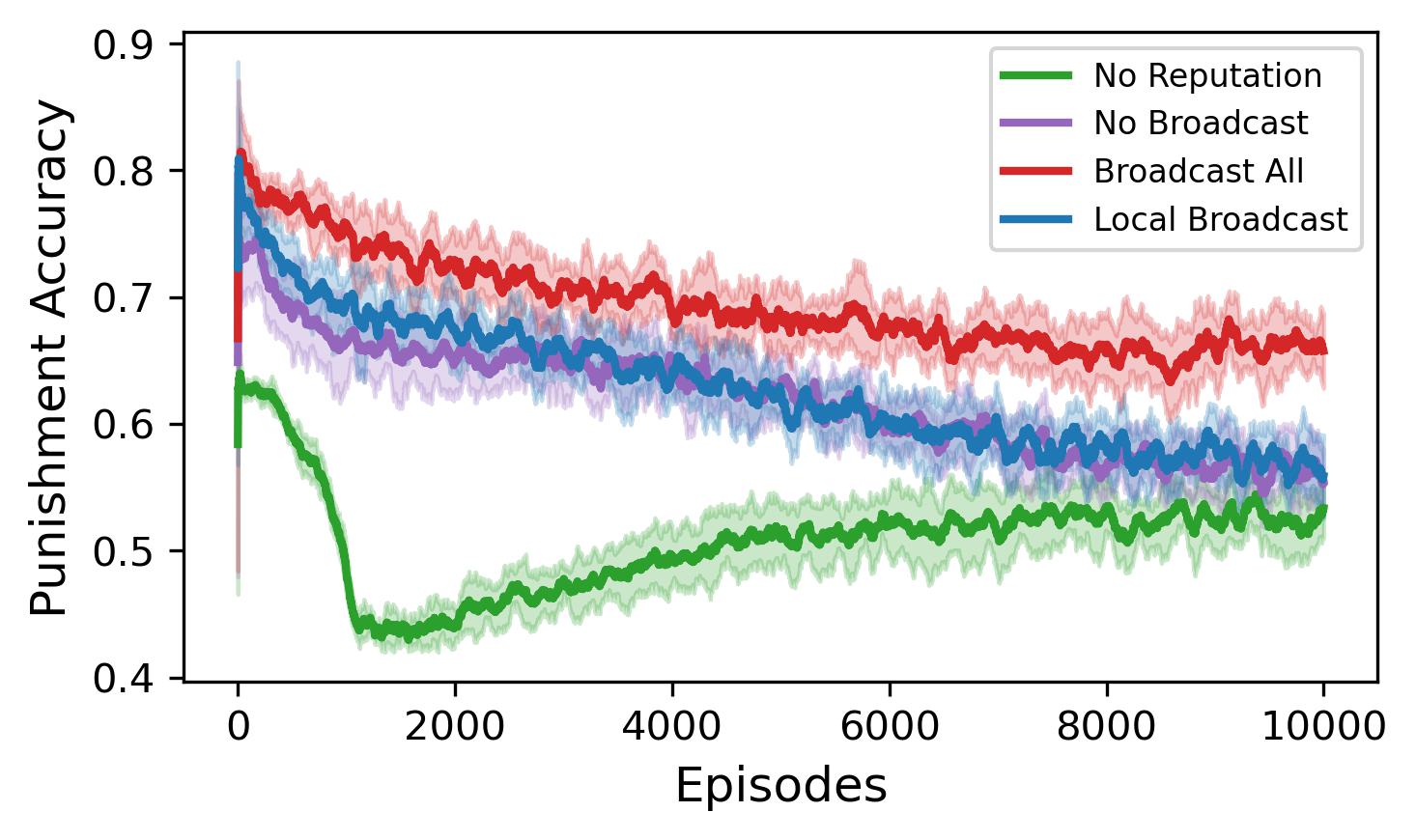}

    \caption{%Punishment Accuracy metric for three reputation methods and baseline throughout trainning. As broadcasting radius increases, agents identifies and tag low reputation agents more accurately.
    Punishment accuracy throughout training for the three reputation methods and the No Reputation baseline on Harvest. Punishment accuracy increases with the broadcast radius, indicating that broader opinion sharing helps agents direct punishment toward low-reputation agents.
    }
    \label{fig::harvest_pa}
\end{figure}

\subsection{Results \& Discussion}
\subsubsection{Prisoner's Dilemma:} %Figure \ref{fig::pd} shows how cooperation evolves along the training process in the Prisoner's Dilemma for three broadcast radii: Broadcast All and Local Broadcast with $\rho = 5$ and $\rho = 10$. The vertical axis is the number of rounds per episode (out of $1000$) in which an agent cooperates, averaged over agents, and the shaded region shows the variance across runs. The results show that reputation with gossip could successfully lead to cooperation compared to the No Reputation baseline, for which players consistently defect.
Figure~\ref{fig::pd} shows how cooperation evolves throughout training in the Prisoner's Dilemma under three broadcasting settings: Broadcast All and Local Broadcast with $\rho = 5$ and $\rho = 10$. Each curve reports the mean across 10 independent runs, with shaded regions indicating the variance across runs. The vertical axis reports the average number of rounds per episode (out of $1000$) in which an agent cooperates. Reputation promotes cooperation across all three broadcasting settings, in contrast to the No Reputation baseline, where agents consistently defect.

%The broadcast radius has little effect on this outcome. The three settings converge to the same level of cooperation, and the differences in when they do so are small relative to the variation across runs: $\rho = 10$ rises marginally earlier than $\rho = 5$, while broadcast-all is the last to reach the plateau and shows the widest variation across runs, both during the transition and afterward. This contrasts with Harvest, where the radius directly affects the outcome. One plausible explanation is that in the Prisoner's Dilemma, every agent directly observes four others in each round, and tables are reshuffled every round, so an agent's first-hand opinions cover the whole population within a few rounds, and gossip adds little that it would not soon observe for itself.

The broadcast radius has little effect on the overall level of cooperation, with all three settings converging to similar outcomes. Differences in convergence speed are small relative to the variation across runs. Local Broadcast with $\rho = 10$ converges slightly faster than with $\rho = 5$, whereas Broadcast All converges more slowly and exhibits the greatest variation, both during training and after convergence. This contrasts with Harvest, where the broadcast radius has a more pronounced effect on collective outcomes. One possible explanation is that, in the Prisoner's Dilemma, each agent directly observes four others per round, and interaction groups are randomly reshuffled. Consequently, agents can accumulate first-hand information about much of the population within relatively few rounds, reducing the additional information provided by gossip.

\subsubsection{Harvest:} %Figures \ref{fig::harvest_u} -- \ref{fig::harvest_pa} present the evolution of utility, peacefulness, equality, sustainability, and punishment accuracy throughout the training process. Each curve is the average of 10 independent runs, and the shaded region shows the variance across runs. The three reputation-based methods differ only in the broadcast radius $\rho$: \textbf{No Broadcast} sets $\rho = 0$, so agents form opinions from first-hand observation alone, \textbf{Local Broadcast} sets $\rho = 5$, so agents exchange opinions only with others nearby, and Broadcast All} sets $\rho = \infty$, so every agent exchanges opinions with the whole population. The \textbf{No Reputation} baseline uses no reputation system. 

Figures~\ref{fig::harvest_u}--\ref{fig::harvest_pa} show the evolution of utility, peacefulness, equality, sustainability, and punishment accuracy throughout training. Each curve reports the mean across 10 independent runs, with shaded regions indicating the variance across runs. The three reputation-based methods differ only in the broadcast radius $\rho$. \textbf{No Broadcast} ($\rho=0$) restricts agents to opinions formed from their own observations; \textbf{Local Broadcast} ($\rho=5$) allows agents to exchange opinions within their local communication neighborhoods; and \textbf{Broadcast All} ($\rho=\infty$) allows opinions to be shared across the entire population. The \textbf{No Reputation} baseline removes the reputation mechanism entirely.

%Figure \ref{fig::harvest_u} shows that all three reputation-based methods substantially outperform the No Reputation baseline in terms of utility, converging to values above 110, whereas the baseline peaks near 67 early in training and then declines to roughly 50 as agents learn to deplete the resource. Among the reputation-based methods, utility increases with the broadcast radius. Broadcast-all is also the slowest to improve early in training, and only overtakes the other two after roughly 5{,}000 episodes.

Figure~\ref{fig::harvest_u} shows that all three reputation-based methods substantially outperform the \textbf{No Reputation} baseline, converging to utility values above 110. In contrast, the baseline peaks near 67 early in training before declining to roughly 50 as agents learn to deplete the shared resource. Among the reputation-based methods, utility increases with the broadcast radius. \textbf{Broadcast All} improves more slowly early in training but eventually surpasses the other two methods after approximately 5000 episodes.

%Sustainability shows the clearest effect of broadcasting (Figure \ref{fig::harvest_s}). The baseline falls steadily from about 480 to 260, meaning that agents collect their rewards earlier and earlier in the episode as they exhaust the apple patches. All reputation-based methods remain well above this level, and sustainability again increases with the broadcast radius. The no-broadcast method, however, exhibits very high variation across runs. Without broadcasting, each agent's opinions rest only on what it has observed first-hand, and whether the population settles on sustainable harvesting differs greatly from one run to the next. In contrast, both broadcasting methods produce tight bands, indicating that sharing opinions makes the sustainable outcome consistent across runs rather than only better on average. Notably, the no-broadcast runs do not differ much in utility, so these runs collect a similar number of apples but at very different times in the episode.

Sustainability exhibits the clearest effect of broadcasting (Figure~\ref{fig::harvest_s}). The No Reputation baseline declines steadily from approximately 480 to 260, indicating that agents collect rewards progressively earlier in the episode as they deplete the apple patches. Reputation-based methods remain substantially above the baseline, with sustainability increasing with the broadcast radius. No Broadcast, however, exhibits considerable variance across runs. One possible explanation is that agents rely exclusively on their own observations, and thus the emergence of sustainable harvesting varies substantially across runs. In contrast, both broadcasting methods exhibit lower variance, suggesting that opinion sharing leads to more consistent sustainable behavior, in addition to improving average sustainability. Interestingly, utility varies much less across No Broadcast runs, indicating that agents collect similar numbers of apples but at substantially different times within episodes.

%Figure \ref{fig::harvest_p} shows that reputation also promotes higher peacefulness, with all three methods exceeding the baseline. Local-broadcast and no-broadcast perform almost identically, while broadcast-all is marginally lower. In terms of equality (Figure \ref{fig::harvest_e}), the no-broadcast method converges to a level comparable to the baseline, whereas the broadcast-all method sacrifices equality in exchange of higher sustainability and utility. We believe this lower equality arises because, when opinions are shared across the whole population, an agent with a poor reputation becomes a target for all other agents and is timed out repeatedly, which prevents it from recovering. This interpretation is consistent with the punishment accuracy results (Figure \ref{fig::harvest_pa}), which measure how often the time-out beam is used on agents with negative reputation. In the broadcast-all method, low-reputation agents are tagged most accurately throughout training, while local-broadcast and no-broadcast converge to a similar level, and the baseline, whose agents are not rewarded on reputation, remains lowest. 

Figure~\ref{fig::harvest_p} shows that reputation also promotes peacefulness, with all reputation-based methods outperforming No Reputation. Local Broadcast and No Broadcast perform similarly, while Broadcast All achieves slightly lower peacefulness. A different pattern emerges for equality (Figure~\ref{fig::harvest_e}). No Broadcast converges to a level comparable to the baseline, whereas Broadcast All achieves higher sustainability and utility at the cost of lower equality. One possible explanation is that population-wide opinion sharing makes an agent with a poor reputation identifiable to all other agents, increasing its exposure to repeated time-outs and potentially limiting its ability to recover its reputation. The punishment accuracy results in Figure~\ref{fig::harvest_pa} are consistent with this interpretation. Broadcast All most consistently directs time-out beams toward agents with negative reputation, whereas Local Broadcast and No Broadcast converge to similar levels of punishment accuracy. The baseline, in which agents do not use reputation, exhibits the lowest punishment accuracy.

%Taken together, these results suggest that local-broadcast is the most balanced of the three settings. The two extremes each have a clear weakness. No-broadcast preserves equality and peacefulness, but it has the lowest sustainability of the reputation-based methods and, more importantly, an outcome that varies widely from run to run. Broadcast-all achieves the highest utility, sustainability, and punishment accuracy, but at the cost of the lowest equality, slightly lower peacefulness, and slower learning early in training. Local-broadcast avoids both weaknesses. It matches no-broadcast on equality and peacefulness, where the two are the best of all methods, while recovering most of the sustainability gained by broadcasting with the same consistency across runs as broadcast-all. Limiting how far opinions travel therefore appears to capture most of the benefit of sharing them, while avoiding the concentrated punishment of low-reputation agents that arises when every opinion reaches the whole population.

Taken together, these results suggest that Local Broadcast provides the most balanced performance across metrics. The two extremes exhibit distinct trade-offs. No Broadcast preserves equality and peacefulness but achieves the lowest sustainability among the reputation-based methods and exhibits substantial variation across runs. Broadcast All achieves the highest utility, sustainability, and punishment accuracy, but with lower equality, slightly lower peacefulness, and slower improvement early in training. Local Broadcast largely avoids these trade-offs: it maintains the high equality and peacefulness of No Broadcast while recovering most of the sustainability benefits and consistency associated with Broadcast All. These results suggest that limiting the reach of opinion sharing can retain much of the benefit of reputation diffusion without the lower equality associated with population-wide broadcasting.
\section{Conclusion \& Future Work}

%In conclusion, we proposed a reputation framework in which agents form private opinions from what they observe, exchange those opinions with their neighbors through DeGroot averaging, and are rewarded on the reputation that results, over a communication network that is redrawn at every step. We studied the framework in an $n$-player Prisoner's Dilemma with randomly regrouped tables and in the Harvest sequential social dilemma, using the broadcast radius to control how far opinions travel. In both games, agents learned to cooperate without prosocial preferences being built into them. 

We introduced a reputation framework in which agents form private opinions from local observations, exchange these opinions with their neighbors through DeGroot averaging, and receive rewards shaped by their resulting reputations as the communication network evolves over time. We studied this framework in an $n$-player Prisoner's Dilemma with randomly regrouped interaction partners and in the Harvest sequential social dilemma, using the broadcast radius to control the reach of opinion sharing. Across both environments, agents learned cooperative strategies without those strategies being explicitly prescribed.

%In the Prisoner's Dilemma, where agents without reputation converged to defection, agents with reputation cooperated in about $95\%$ of rounds at every broadcast radius we tested. The radius had little effect on the average level of cooperation, but broadcasting to the whole population made the outcome vary more from run to run.

In the Prisoner's Dilemma, agents without reputation converged to defection, whereas agents with reputation cooperated in approximately $95\%$ of rounds across all broadcast radii. The broadcast radius had little effect on average cooperation, although population-wide broadcasting produced greater variation across runs.

%In Harvest, reputation more than doubled utility relative to a baseline without reputation and produced more peaceful and more sustainable episodes, and a larger radius raised sustainability further and made it consistent across runs. However, more communication was not uniformly better: broadcasting to the whole population gave the highest utility and sustainability but the lowest equality, while broadcasting only to nearby agents retained most of the benefit without that cost. These findings suggest that how far reputation travels is itself a design choice that shapes not only whether agents cooperate, but how the gains from cooperation are distributed.

In Harvest, reputation more than doubled utility relative to the baseline while also promoting more peaceful and sustainable behavior. Increasing the broadcast radius further improved sustainability and reduced variance across runs. However, broader opinion sharing was not uniformly beneficial: population-wide broadcasting achieved the highest utility and sustainability but the lowest equality, whereas local broadcasting retained most of these benefits while preserving equality and peacefulness. These findings show that the reach of reputation diffusion is itself an important design choice, shaping not only cooperation and collective performance but also how their benefits are distributed across agents.

%One limitation of our future work is that it fixes two components that other work has shown can be learned. Agents assess one another with a hand-specified rule and combine what they hear by simple averaging, whereas LR2~\cite{LR2} and COOPER~\cite{cooper} learn how agents judge and, in COOPER's case, how they weigh what their neighbors report. We made this choice so that the effect of the broadcast radius could be isolated, but it leaves open whether learned assessment and learned aggregation remain effective when the communication network changes at every step. Combining the two is a natural next step. A related assumption is that agents report their opinions truthfully. An agent that could choose what to share might spread false assessments to damage a rival or protect itself, and studying gossip that is strategic or noisy would test how far our results depend on honest communication.

One limitation of our work is that we fix two components that prior work has shown can be learned. Agents assess one another using fixed rules and aggregate received opinions through simple averaging, whereas LR2~\cite{LR2} and COOPER~\cite{cooper} learn assessment rules and, in the case of COOPER, how agents weigh information received from their neighbors. We deliberately fix these components to isolate the effect of reputation diffusion, but this leaves open whether learned assessment and aggregation remain effective as communication networks evolve. Extending our framework to jointly learn these mechanisms is a natural direction for future work. We also assume that agents communicate their opinions truthfully. Strategic agents could instead manipulate the information they share, for example, by spreading false assessments of competitors or concealing unfavorable information about themselves. Studying strategic or noisy gossip would therefore help establish how robust our findings are to imperfect communication.
%%%%%%%%%%%%%%%%%%%%%%%%%%%%%%%%%%%%%%%%%%%%%%%%%%%%%%%%%%%%%%%%%%%%%%%%

%%% The next two lines define, first, the bibliography style to be 
%%% applied, and, second, the bibliography file to be used.

\bibliographystyle{ACM-Reference-Format} 
\bibliography{sample}


\begin{thebibliography}{53}

%%% the Web address.

\ifx \showCODEN    \undefined \def \showCODEN     #1{\unskip}     \fi
\ifx \showISBNx    \undefined \def \showISBNx     #1{\unskip}     \fi
\ifx \showISBNxiii \undefined \def \showISBNxiii  #1{\unskip}     \fi
\ifx \showISSN     \undefined \def \showISSN      #1{\unskip}     \fi
\ifx \showLCCN     \undefined \def \showLCCN      #1{\unskip}     \fi
\ifx \shownote     \undefined \def \shownote      #1{#1}          \fi
\ifx \showarticletitle \undefined \def \showarticletitle #1{#1}   \fi
\ifx \showURL      \undefined \def \showURL       {\relax}        \fi
% invisible to TeX
\providecommand\bibfield[2]{#2}
\providecommand\bibinfo[2]{#2}
\providecommand\natexlab[1]{#1}
\providecommand\showeprint[2][]{arXiv:#2}

\bibitem[Anastassacos et~al\mbox{.}(2021)]%
        {anastassacos_21}
\bibfield{author}{\bibinfo{person}{Nicolas Anastassacos}, \bibinfo{person}{Julian Garcia}, \bibinfo{person}{Stephen Hailes}, \bibinfo{person}{Mirco Musolesi}, {et~al\mbox{.}}} \bibinfo{year}{2021}\natexlab{}.
\newblock \showarticletitle{Cooperation and Reputation Dynamics with Reinforcement Learning}. In \bibinfo{booktitle}{\emph{Proceedings of the 20th International Conference on Autonomous Agents and Multiagent Systems (AAMAS 2021)}}. \bibinfo{pages}{115--123}.
\newblock


\bibitem[Baker(2020)]%
        {baker_20}
\bibfield{author}{\bibinfo{person}{Bowen Baker}.} \bibinfo{year}{2020}\natexlab{}.
\newblock \showarticletitle{Emergent reciprocity and team formation from randomized uncertain social preferences}.
\newblock \bibinfo{journal}{\emph{Advances in neural information processing systems}}  \bibinfo{volume}{33} (\bibinfo{year}{2020}), \bibinfo{pages}{15786--15799}.
\newblock


\bibitem[Christoffersen et~al\mbox{.}(2023)]%
        {christoffersen_23}
\bibfield{author}{\bibinfo{person}{Phillip~J.K. Christoffersen}, \bibinfo{person}{Andreas~A. Haupt}, {and} \bibinfo{person}{Dylan Hadfield-Menell}.} \bibinfo{year}{2023}\natexlab{}.
\newblock \showarticletitle{Get It in Writing: Formal Contracts Mitigate Social Dilemmas in Multi-Agent RL}. In \bibinfo{booktitle}{\emph{Proceedings of the 2023 International Conference on Autonomous Agents and Multiagent Systems (AAMAS 2023)}}. \bibinfo{pages}{448–456}.
\newblock
\showISBNx{9781450394321}


\bibitem[Cushman(2008)]%
        {cushman_08}
\bibfield{author}{\bibinfo{person}{Fiery Cushman}.} \bibinfo{year}{2008}\natexlab{}.
\newblock \showarticletitle{Crime and punishment: Distinguishing the roles of causal and intentional analyses in moral judgment}.
\newblock \bibinfo{journal}{\emph{Cognition}} \bibinfo{volume}{108}, \bibinfo{number}{2} (\bibinfo{year}{2008}), \bibinfo{pages}{353--380}.
\newblock


\bibitem[DeGroot(1974)]%
        {DeGroot_74}
\bibfield{author}{\bibinfo{person}{Morris~H DeGroot}.} \bibinfo{year}{1974}\natexlab{}.
\newblock \showarticletitle{Reaching a consensus}.
\newblock \bibinfo{journal}{\emph{J. Amer. Statist. Assoc.}} \bibinfo{volume}{69}, \bibinfo{number}{345} (\bibinfo{year}{1974}), \bibinfo{pages}{118--121}.
\newblock


\bibitem[Dellarocas(2003)]%
        {dellarocas_03}
\bibfield{author}{\bibinfo{person}{Chrysanthos Dellarocas}.} \bibinfo{year}{2003}\natexlab{}.
\newblock \showarticletitle{The digitization of word of mouth: Promise and challenges of online feedback mechanisms}.
\newblock \bibinfo{journal}{\emph{Management science}} \bibinfo{volume}{49}, \bibinfo{number}{10} (\bibinfo{year}{2003}), \bibinfo{pages}{1407--1424}.
\newblock


\bibitem[DeMarzo et~al\mbox{.}(2003)]%
        {DeMarzo_03}
\bibfield{author}{\bibinfo{person}{Peter~M. DeMarzo}, \bibinfo{person}{Dimitri Vayanos}, {and} \bibinfo{person}{Jeffrey Zwiebel}.} \bibinfo{year}{2003}\natexlab{}.
\newblock \showarticletitle{Persuasion Bias, Social Influence, and Unidimensional Opinions}.
\newblock \bibinfo{journal}{\emph{The Quarterly Journal of Economics}} \bibinfo{volume}{118}, \bibinfo{number}{3} (\bibinfo{date}{08} \bibinfo{year}{2003}), \bibinfo{pages}{909--968}.
\newblock


\bibitem[Dietz et~al\mbox{.}(2003)]%
        {dietz_03}
\bibfield{author}{\bibinfo{person}{Thomas Dietz}, \bibinfo{person}{Elinor Ostrom}, {and} \bibinfo{person}{Paul~C Stern}.} \bibinfo{year}{2003}\natexlab{}.
\newblock \showarticletitle{The struggle to govern the commons}.
\newblock \bibinfo{journal}{\emph{Science}} \bibinfo{volume}{302}, \bibinfo{number}{5652} (\bibinfo{year}{2003}), \bibinfo{pages}{1907--1912}.
\newblock


\bibitem[Dores~Cruz et~al\mbox{.}(2021)]%
        {cruz_21}
\bibfield{author}{\bibinfo{person}{Terence~D Dores~Cruz}, \bibinfo{person}{Isabel Thielmann}, \bibinfo{person}{Simon Columbus}, \bibinfo{person}{Catherine Molho}, \bibinfo{person}{Junhui Wu}, \bibinfo{person}{Francesca Righetti}, \bibinfo{person}{Reinout~E De~Vries}, \bibinfo{person}{Antonis Koutsoumpis}, \bibinfo{person}{Paul~AM Van~Lange}, \bibinfo{person}{Bianca Beersma}, {et~al\mbox{.}}} \bibinfo{year}{2021}\natexlab{}.
\newblock \showarticletitle{Gossip and reputation in everyday life}.
\newblock \bibinfo{journal}{\emph{Philosophical Transactions of the Royal Society B: Biological Sciences}} \bibinfo{volume}{376}, \bibinfo{number}{1838} (\bibinfo{year}{2021}).
\newblock


\bibitem[Fatima et~al\mbox{.}(2024)]%
        {fatima_24}
\bibfield{author}{\bibinfo{person}{Shaheen Fatima}, \bibinfo{person}{Nicholas~R Jennings}, {and} \bibinfo{person}{Michael Wooldridge}.} \bibinfo{year}{2024}\natexlab{}.
\newblock \showarticletitle{Learning to resolve social dilemmas: a survey}.
\newblock \bibinfo{journal}{\emph{Journal of Artificial Intelligence Research}}  \bibinfo{volume}{79} (\bibinfo{year}{2024}), \bibinfo{pages}{895--969}.
\newblock


\bibitem[Fehr and Fischbacher(2004)]%
        {fehr_04}
\bibfield{author}{\bibinfo{person}{Ernst Fehr} {and} \bibinfo{person}{Urs Fischbacher}.} \bibinfo{year}{2004}\natexlab{}.
\newblock \showarticletitle{Third-party punishment and social norms}.
\newblock \bibinfo{journal}{\emph{Evolution and Human Behavior}} \bibinfo{volume}{25}, \bibinfo{number}{2} (\bibinfo{year}{2004}), \bibinfo{pages}{63--87}.
\newblock
\showISSN{1090-5138}
\href{https://doi.org/10.1016/S1090-5138(04)00005-4}{doi:\nolinkurl{10.1016/S1090-5138(04)00005-4}}


\bibitem[Fehr and G{\"a}chter(2002)]%
        {fehr_02}
\bibfield{author}{\bibinfo{person}{Ernst Fehr} {and} \bibinfo{person}{Simon G{\"a}chter}.} \bibinfo{year}{2002}\natexlab{}.
\newblock \showarticletitle{Altruistic punishment in humans}.
\newblock \bibinfo{journal}{\emph{Nature}} \bibinfo{volume}{415}, \bibinfo{number}{6868} (\bibinfo{year}{2002}), \bibinfo{pages}{137--140}.
\newblock


\bibitem[Feinberg et~al\mbox{.}(2014)]%
        {feinberg_14}
\bibfield{author}{\bibinfo{person}{Matthew Feinberg}, \bibinfo{person}{Robb Willer}, {and} \bibinfo{person}{Michael Schultz}.} \bibinfo{year}{2014}\natexlab{}.
\newblock \showarticletitle{Gossip and ostracism promote cooperation in groups}.
\newblock \bibinfo{journal}{\emph{Psychological science}} \bibinfo{volume}{25}, \bibinfo{number}{3} (\bibinfo{year}{2014}), \bibinfo{pages}{656--664}.
\newblock


\bibitem[Foerster et~al\mbox{.}(2018)]%
        {Foerster_18}
\bibfield{author}{\bibinfo{person}{Jakob Foerster}, \bibinfo{person}{Richard~Y. Chen}, \bibinfo{person}{Maruan Al-Shedivat}, \bibinfo{person}{Shimon Whiteson}, \bibinfo{person}{Pieter Abbeel}, {and} \bibinfo{person}{Igor Mordatch}.} \bibinfo{year}{2018}\natexlab{}.
\newblock \showarticletitle{Learning with opponent-learning awareness}. In \bibinfo{booktitle}{\emph{Autonomous Agents and Multi-Agent Systems (AAMAS 2018)}}.
\newblock


\bibitem[Giardini et~al\mbox{.}(2014)]%
        {giardini_14}
\bibfield{author}{\bibinfo{person}{Francesca Giardini}, \bibinfo{person}{Mario Paolucci}, \bibinfo{person}{Daniel Villatoro}, {and} \bibinfo{person}{Rosaria Conte}.} \bibinfo{year}{2014}\natexlab{}.
\newblock \showarticletitle{Punishment and gossip: sustaining cooperation in a public goods game}. In \bibinfo{booktitle}{\emph{Advances in Social Simulation: Proceedings of the 9th Conference of the European Social Simulation Association}}. Springer, \bibinfo{pages}{107--118}.
\newblock


\bibitem[Golub and Jackson(2010)]%
        {golub_10}
\bibfield{author}{\bibinfo{person}{Benjamin Golub} {and} \bibinfo{person}{Matthew~O Jackson}.} \bibinfo{year}{2010}\natexlab{}.
\newblock \showarticletitle{Naive learning in social networks and the wisdom of crowds}.
\newblock \bibinfo{journal}{\emph{American Economic Journal: Microeconomics}} \bibinfo{volume}{2}, \bibinfo{number}{1} (\bibinfo{year}{2010}), \bibinfo{pages}{112--149}.
\newblock


\bibitem[Hardin(1968)]%
        {hardin_68}
\bibfield{author}{\bibinfo{person}{Garrett Hardin}.} \bibinfo{year}{1968}\natexlab{}.
\newblock \showarticletitle{The tragedy of the commons: the population problem has no technical solution; it requires a fundamental extension in morality.}
\newblock \bibinfo{journal}{\emph{Science}} \bibinfo{volume}{162}, \bibinfo{number}{3859} (\bibinfo{year}{1968}), \bibinfo{pages}{1243--1248}.
\newblock


\bibitem[Hilbe et~al\mbox{.}(2018)]%
        {Hilbe_18}
\bibfield{author}{\bibinfo{person}{Christian Hilbe}, \bibinfo{person}{Laura Schmid}, \bibinfo{person}{Josef Tkadlec}, \bibinfo{person}{Krishnendu Chatterjee}, {and} \bibinfo{person}{Martin~A Nowak}.} \bibinfo{year}{2018}\natexlab{}.
\newblock \showarticletitle{Indirect reciprocity with private, noisy, and incomplete information}.
\newblock \bibinfo{journal}{\emph{Proceedings of the National Academy of Sciences}} \bibinfo{volume}{115}, \bibinfo{number}{48} (\bibinfo{year}{2018}), \bibinfo{pages}{12241--12246}.
\newblock


\bibitem[Hughes et~al\mbox{.}(2018)]%
        {hughes_18}
\bibfield{author}{\bibinfo{person}{Edward Hughes}, \bibinfo{person}{Joel~Z Leibo}, \bibinfo{person}{Matthew Phillips}, \bibinfo{person}{Karl Tuyls}, \bibinfo{person}{Edgar Due{\~n}ez-Guzman}, \bibinfo{person}{Antonio Garc{\'\i}a~Casta{\~n}eda}, \bibinfo{person}{Iain Dunning}, \bibinfo{person}{Tina Zhu}, \bibinfo{person}{Kevin McKee}, \bibinfo{person}{Raphael Koster}, {et~al\mbox{.}}} \bibinfo{year}{2018}\natexlab{}.
\newblock \showarticletitle{Inequity aversion improves cooperation in intertemporal social dilemmas}.
\newblock \bibinfo{journal}{\emph{Advances in Neural Information Processing Systems}}  \bibinfo{volume}{31} (\bibinfo{year}{2018}).
\newblock


\bibitem[Kawakatsu et~al\mbox{.}(2024)]%
        {Kawakatsu_24}
\bibfield{author}{\bibinfo{person}{Mari Kawakatsu}, \bibinfo{person}{Taylor~A Kessinger}, {and} \bibinfo{person}{Joshua~B Plotkin}.} \bibinfo{year}{2024}\natexlab{}.
\newblock \showarticletitle{A mechanistic model of gossip, reputations, and cooperation}.
\newblock \bibinfo{journal}{\emph{Proceedings of the National Academy of Sciences}} \bibinfo{volume}{121}, \bibinfo{number}{20} (\bibinfo{year}{2024}), \bibinfo{pages}{e2400689121}.
\newblock


\bibitem[Leibo et~al\mbox{.}(2017)]%
        {leibo_17_multi}
\bibfield{author}{\bibinfo{person}{Joel~Z Leibo}, \bibinfo{person}{Vinicius Zambaldi}, \bibinfo{person}{Marc Lanctot}, \bibinfo{person}{Janusz Marecki}, {and} \bibinfo{person}{Thore Graepel}.} \bibinfo{year}{2017}\natexlab{}.
\newblock \showarticletitle{Multi-agent Reinforcement Learning in Sequential Social Dilemmas}. In \bibinfo{booktitle}{\emph{Proceedings of the 16th Conference on Autonomous Agents and MultiAgent Systems}}. \bibinfo{pages}{464--473}.
\newblock


\bibitem[Lerer and Peysakhovich(2018)]%
        {lerer_18}
\bibfield{author}{\bibinfo{person}{Adam Lerer} {and} \bibinfo{person}{Alexander Peysakhovich}.} \bibinfo{year}{2018}\natexlab{}.
\newblock \bibinfo{title}{Maintaining cooperation in complex social dilemmas using deep reinforcement learning}.
\newblock
\showeprint[arxiv]{1707.01068}~[cs.AI]


\bibitem[Lu et~al\mbox{.}(2022)]%
        {lu_22}
\bibfield{author}{\bibinfo{person}{Christopher Lu}, \bibinfo{person}{Timon Willi}, \bibinfo{person}{Christian A~Schroeder De~Witt}, {and} \bibinfo{person}{Jakob Foerster}.} \bibinfo{year}{2022}\natexlab{}.
\newblock \showarticletitle{Model-free opponent shaping}. In \bibinfo{booktitle}{\emph{International Conference on Machine Learning}}. PMLR, \bibinfo{pages}{14398--14411}.
\newblock


\bibitem[Lupu and Precup(2020)]%
        {Lupu_20}
\bibfield{author}{\bibinfo{person}{Andrei Lupu} {and} \bibinfo{person}{Doina Precup}.} \bibinfo{year}{2020}\natexlab{}.
\newblock \showarticletitle{Gifting in multi-agent reinforcement learning}. In \bibinfo{booktitle}{\emph{Proceedings of the 19th International Conference on Autonomous Agents and Multiagent Systems}}. \bibinfo{pages}{789--797}.
\newblock


\bibitem[McKee et~al\mbox{.}(2020)]%
        {mckee_20}
\bibfield{author}{\bibinfo{person}{Kevin~R McKee}, \bibinfo{person}{Ian Gemp}, \bibinfo{person}{Brian McWilliams}, \bibinfo{person}{Edgar~A Du{\`e}{\~n}ez-Guzm{\'a}n}, \bibinfo{person}{Edward Hughes}, {and} \bibinfo{person}{Joel~Z Leibo}.} \bibinfo{year}{2020}\natexlab{}.
\newblock \showarticletitle{Social Diversity and Social Preferences in Mixed-Motive Reinforcement Learning}. In \bibinfo{booktitle}{\emph{Proceedings of the 19th International Conference on Autonomous Agents and MultiAgent Systems}}. \bibinfo{pages}{869--877}.
\newblock


\bibitem[McKee et~al\mbox{.}(2021)]%
        {mckee_23}
\bibfield{author}{\bibinfo{person}{Kevin~R McKee}, \bibinfo{person}{Edward Hughes}, \bibinfo{person}{Tina~O Zhu}, \bibinfo{person}{Martin~J Chadwick}, \bibinfo{person}{Raphael Koster}, \bibinfo{person}{Antonio~Garcia Castaneda}, \bibinfo{person}{Charlie Beattie}, \bibinfo{person}{Thore Graepel}, \bibinfo{person}{Matt Botvinick}, {and} \bibinfo{person}{Joel~Z Leibo}.} \bibinfo{year}{2021}\natexlab{}.
\newblock \showarticletitle{A multi-agent reinforcement learning model of reputation and cooperation in human groups}.
\newblock \bibinfo{journal}{\emph{arXiv preprint arXiv:2103.04982}} (\bibinfo{year}{2021}).
\newblock


\bibitem[Milinski et~al\mbox{.}(2002)]%
        {milinski_02}
\bibfield{author}{\bibinfo{person}{Manfred Milinski}, \bibinfo{person}{Dirk Semmann}, {and} \bibinfo{person}{Hans-J{\"u}rgen Krambeck}.} \bibinfo{year}{2002}\natexlab{}.
\newblock \showarticletitle{Reputation helps solve the ‘tragedy of the commons’}.
\newblock \bibinfo{journal}{\emph{Nature}} \bibinfo{volume}{415}, \bibinfo{number}{6870} (\bibinfo{year}{2002}), \bibinfo{pages}{424--426}.
\newblock


\bibitem[Mnih et~al\mbox{.}(2015)]%
        {DQN_15}
\bibfield{author}{\bibinfo{person}{Volodymyr Mnih}, \bibinfo{person}{Koray Kavukcuoglu}, \bibinfo{person}{David Silver}, \bibinfo{person}{Andrei~A Rusu}, \bibinfo{person}{Joel Veness}, \bibinfo{person}{Marc~G Bellemare}, \bibinfo{person}{Alex Graves}, \bibinfo{person}{Martin Riedmiller}, \bibinfo{person}{Andreas~K Fidjeland}, \bibinfo{person}{Georg Ostrovski}, {et~al\mbox{.}}} \bibinfo{year}{2015}\natexlab{}.
\newblock \showarticletitle{Human-level control through deep reinforcement learning}.
\newblock \bibinfo{journal}{\emph{Nature}} \bibinfo{volume}{518}, \bibinfo{number}{7540} (\bibinfo{year}{2015}), \bibinfo{pages}{529--533}.
\newblock


\bibitem[Molho and Wu(2021)]%
        {molho_21}
\bibfield{author}{\bibinfo{person}{Catherine Molho} {and} \bibinfo{person}{Junhui Wu}.} \bibinfo{year}{2021}\natexlab{}.
\newblock \showarticletitle{Direct punishment and indirect reputation-based tactics to intervene against offences}.
\newblock \bibinfo{journal}{\emph{Philosophical Transactions of the Royal Society B: Biological Sciences}} \bibinfo{volume}{376}, \bibinfo{number}{1838} (\bibinfo{year}{2021}).
\newblock


\bibitem[Murase and Hilbe(2023)]%
        {murase_23}
\bibfield{author}{\bibinfo{person}{Yohsuke Murase} {and} \bibinfo{person}{Christian Hilbe}.} \bibinfo{year}{2023}\natexlab{}.
\newblock \showarticletitle{Indirect reciprocity with stochastic and dual reputation updates}.
\newblock \bibinfo{journal}{\emph{PLOS Computational Biology}} \bibinfo{volume}{19}, \bibinfo{number}{7} (\bibinfo{year}{2023}), \bibinfo{pages}{e1011271}.
\newblock


\bibitem[Murase and Hilbe(2024)]%
        {murase_24}
\bibfield{author}{\bibinfo{person}{Yohsuke Murase} {and} \bibinfo{person}{Christian Hilbe}.} \bibinfo{year}{2024}\natexlab{}.
\newblock \showarticletitle{Indirect reciprocity under opinion synchronization}.
\newblock \bibinfo{journal}{\emph{Proceedings of the National Academy of Sciences}} \bibinfo{volume}{121}, \bibinfo{number}{48} (\bibinfo{year}{2024}).
\newblock


\bibitem[Nowak(2006)]%
        {nowak_06}
\bibfield{author}{\bibinfo{person}{Martin~A. Nowak}.} \bibinfo{year}{2006}\natexlab{}.
\newblock \showarticletitle{Five rules for the evolution of cooperation}.
\newblock \bibinfo{journal}{\emph{Science}} \bibinfo{volume}{314}, \bibinfo{number}{5805} (\bibinfo{date}{dec} \bibinfo{year}{2006}), \bibinfo{pages}{1560--1563}.
\newblock
\href{https://doi.org/10.1126/science.1133755}{doi:\nolinkurl{10.1126/science.1133755}}


\bibitem[Nowak and Sigmund(1998)]%
        {nowak_98}
\bibfield{author}{\bibinfo{person}{Martin~A Nowak} {and} \bibinfo{person}{Karl Sigmund}.} \bibinfo{year}{1998}\natexlab{}.
\newblock \showarticletitle{Evolution of indirect reciprocity by image scoring}.
\newblock \bibinfo{journal}{\emph{Nature}} \bibinfo{volume}{393}, \bibinfo{number}{6685} (\bibinfo{year}{1998}), \bibinfo{pages}{573--577}.
\newblock


\bibitem[Ohtsuki and Iwasa(2006)]%
        {ohtsuki_06}
\bibfield{author}{\bibinfo{person}{Hisashi Ohtsuki} {and} \bibinfo{person}{Yoh Iwasa}.} \bibinfo{year}{2006}\natexlab{}.
\newblock \showarticletitle{The leading eight: social norms that can maintain cooperation by indirect reciprocity}.
\newblock \bibinfo{journal}{\emph{Journal of theoretical biology}} \bibinfo{volume}{239}, \bibinfo{number}{4} (\bibinfo{year}{2006}), \bibinfo{pages}{435--444}.
\newblock


\bibitem[Oroojlooy and Hajinezhad(2023)]%
        {Oroojlooy_22}
\bibfield{author}{\bibinfo{person}{Afshin Oroojlooy} {and} \bibinfo{person}{Davood Hajinezhad}.} \bibinfo{year}{2023}\natexlab{}.
\newblock \showarticletitle{A review of cooperative multi-agent deep reinforcement learning}.
\newblock \bibinfo{journal}{\emph{Applied Intelligence}} \bibinfo{volume}{53}, \bibinfo{number}{11} (\bibinfo{year}{2023}), \bibinfo{pages}{13677--13722}.
\newblock


\bibitem[Orzan et~al\mbox{.}(2024)]%
        {orzan_24}
\bibfield{author}{\bibinfo{person}{Nicole Orzan}, \bibinfo{person}{Erman Acar}, \bibinfo{person}{Davide Grossi}, {and} \bibinfo{person}{Roxana Radulescu}.} \bibinfo{year}{2024}\natexlab{}.
\newblock \showarticletitle{Emergent Cooperation under Uncertain Incentive Alignment}. In \bibinfo{booktitle}{\emph{The 23rd International Conference on Autonomous Agents and Multi-Agent Systems}}. International Foundation for Autonomous Agents and Multiagent Systems (IFAAMAS), \bibinfo{pages}{1521--1530}.
\newblock


\bibitem[Ostrom et~al\mbox{.}(1990)]%
        {ostrom_90}
\bibfield{author}{\bibinfo{person}{Elinor Ostrom} {et~al\mbox{.}}} \bibinfo{year}{1990}\natexlab{}.
\newblock \bibinfo{booktitle}{\emph{Governing the commons: The evolution of institutions for collective action}}. Vol.~\bibinfo{volume}{10}.
\newblock \bibinfo{publisher}{HeinOnline}.
\newblock


\bibitem[Ostrom et~al\mbox{.}(1999)]%
        {ostrom_99}
\bibfield{author}{\bibinfo{person}{Elinor Ostrom}, \bibinfo{person}{Joanna Burger}, \bibinfo{person}{Christopher~B Field}, \bibinfo{person}{Richard~B Norgaard}, {and} \bibinfo{person}{David Policansky}.} \bibinfo{year}{1999}\natexlab{}.
\newblock \showarticletitle{Revisiting the commons: local lessons, global challenges}.
\newblock \bibinfo{journal}{\emph{Science}} \bibinfo{volume}{284}, \bibinfo{number}{5412} (\bibinfo{year}{1999}), \bibinfo{pages}{278--282}.
\newblock


\bibitem[Pacheco et~al\mbox{.}(2006)]%
        {Pacheco_06}
\bibfield{author}{\bibinfo{person}{Jorge~M Pacheco}, \bibinfo{person}{Arne Traulsen}, {and} \bibinfo{person}{Martin~A Nowak}.} \bibinfo{year}{2006}\natexlab{}.
\newblock \showarticletitle{Coevolution of strategy and structure in complex networks with dynamical linking}.
\newblock \bibinfo{journal}{\emph{Physical review letters}} \bibinfo{volume}{97}, \bibinfo{number}{25} (\bibinfo{year}{2006}), \bibinfo{pages}{258103}.
\newblock


\bibitem[Pan et~al\mbox{.}(2024)]%
        {pan_24}
\bibfield{author}{\bibinfo{person}{Xinyue Pan}, \bibinfo{person}{Vincent Hsiao}, \bibinfo{person}{Dana~S Nau}, {and} \bibinfo{person}{Michele~J Gelfand}.} \bibinfo{year}{2024}\natexlab{}.
\newblock \showarticletitle{Explaining the evolution of gossip}.
\newblock \bibinfo{journal}{\emph{Proceedings of the National Academy of Sciences}} \bibinfo{volume}{121}, \bibinfo{number}{9} (\bibinfo{year}{2024}), \bibinfo{pages}{e2214160121}.
\newblock


\bibitem[Pennisi(2005)]%
        {Pennisi_05}
\bibfield{author}{\bibinfo{person}{Elizabeth Pennisi}.} \bibinfo{year}{2005}\natexlab{}.
\newblock \showarticletitle{How did cooperative behavior evolve?}
\newblock \bibinfo{journal}{\emph{Science}} \bibinfo{volume}{309}, \bibinfo{number}{5731} (\bibinfo{year}{2005}), \bibinfo{pages}{93--93}.
\newblock


\bibitem[Perolat et~al\mbox{.}(2017)]%
        {Leibo_17}
\bibfield{author}{\bibinfo{person}{Julien Perolat}, \bibinfo{person}{Joel~Z Leibo}, \bibinfo{person}{Vinicius Zambaldi}, \bibinfo{person}{Charles Beattie}, \bibinfo{person}{Karl Tuyls}, {and} \bibinfo{person}{Thore Graepel}.} \bibinfo{year}{2017}\natexlab{}.
\newblock \showarticletitle{A multi-agent reinforcement learning model of common-pool resource appropriation}.
\newblock \bibinfo{journal}{\emph{Advances in neural information processing systems}}  \bibinfo{volume}{30} (\bibinfo{year}{2017}).
\newblock


\bibitem[Peysakhovich and Lerer(2018)]%
        {peysakhovich_18}
\bibfield{author}{\bibinfo{person}{Alexander Peysakhovich} {and} \bibinfo{person}{Adam Lerer}.} \bibinfo{year}{2018}\natexlab{}.
\newblock \showarticletitle{Consequentialist conditional cooperation in social dilemmas with imperfect information}. In \bibinfo{booktitle}{\emph{International Conference on Learning Representations}}.
\newblock


\bibitem[Rand et~al\mbox{.}(2011)]%
        {Rand_11}
\bibfield{author}{\bibinfo{person}{David~G Rand}, \bibinfo{person}{Samuel Arbesman}, {and} \bibinfo{person}{Nicholas~A Christakis}.} \bibinfo{year}{2011}\natexlab{}.
\newblock \showarticletitle{Dynamic social networks promote cooperation in experiments with humans}.
\newblock \bibinfo{journal}{\emph{Proceedings of the National Academy of Sciences}} \bibinfo{volume}{108}, \bibinfo{number}{48} (\bibinfo{year}{2011}), \bibinfo{pages}{19193--19198}.
\newblock


\bibitem[Ren et~al\mbox{.}(2025)]%
        {LR2}
\bibfield{author}{\bibinfo{person}{Tianyu Ren}, \bibinfo{person}{Xuan Yao}, \bibinfo{person}{Yang Li}, {and} \bibinfo{person}{Xiao-Jun Zeng}.} \bibinfo{year}{2025}\natexlab{}.
\newblock \showarticletitle{Bottom-Up Reputation Promotes Cooperation with Multi-Agent Reinforcement Learning}. In \bibinfo{booktitle}{\emph{Proceedings of the 24th International Conference on Autonomous Agents and Multiagent Systems}}. \bibinfo{pages}{1745--1754}.
\newblock


\bibitem[Sitthiyot and Holasut(2020)]%
        {Sitthiyot_20}
\bibfield{author}{\bibinfo{person}{Thitithep Sitthiyot} {and} \bibinfo{person}{Kanyarat Holasut}.} \bibinfo{year}{2020}\natexlab{}.
\newblock \showarticletitle{A simple method for measuring inequality}.
\newblock \bibinfo{journal}{\emph{Palgrave Communications}} \bibinfo{volume}{6}, \bibinfo{number}{1} (\bibinfo{year}{2020}), \bibinfo{pages}{112}.
\newblock


\bibitem[Sommerfeld et~al\mbox{.}(2007)]%
        {sommerfeld_07}
\bibfield{author}{\bibinfo{person}{Ralf~D Sommerfeld}, \bibinfo{person}{Hans-J{\"u}rgen Krambeck}, \bibinfo{person}{Dirk Semmann}, {and} \bibinfo{person}{Manfred Milinski}.} \bibinfo{year}{2007}\natexlab{}.
\newblock \showarticletitle{Gossip as an alternative for direct observation in games of indirect reciprocity}.
\newblock \bibinfo{journal}{\emph{Proceedings of the national academy of sciences}} \bibinfo{volume}{104}, \bibinfo{number}{44} (\bibinfo{year}{2007}), \bibinfo{pages}{17435--17440}.
\newblock


\bibitem[Song et~al\mbox{.}(2026)]%
        {cooper}
\bibfield{author}{\bibinfo{person}{Xinwei Song}, \bibinfo{person}{Yizhe Huang}, \bibinfo{person}{Dengji Zhao}, {and} \bibinfo{person}{Xue Feng}.} \bibinfo{year}{2026}\natexlab{}.
\newblock \showarticletitle{Learning to cooperate with emergent reputation via multi-agent reinforcement learning}.
\newblock \bibinfo{journal}{\emph{arXiv preprint arXiv:2606.04359}} (\bibinfo{year}{2026}).
\newblock


\bibitem[Su et~al\mbox{.}(2023)]%
        {Su_23}
\bibfield{author}{\bibinfo{person}{Qi Su}, \bibinfo{person}{Alex McAvoy}, {and} \bibinfo{person}{Joshua~B Plotkin}.} \bibinfo{year}{2023}\natexlab{}.
\newblock \showarticletitle{Strategy evolution on dynamic networks}.
\newblock \bibinfo{journal}{\emph{Nature Computational Science}} \bibinfo{volume}{3}, \bibinfo{number}{9} (\bibinfo{year}{2023}), \bibinfo{pages}{763--776}.
\newblock


\bibitem[Vilone et~al\mbox{.}({[n.\,d.]})]%
        {vilone_16}
\bibfield{author}{\bibinfo{person}{Daniele Vilone}, \bibinfo{person}{Francesca Giardini}, {and} \bibinfo{person}{Mario Paolucci}.} \bibinfo{year}{[n.\,d.]}\natexlab{}.
\newblock \showarticletitle{Exploring Reputation-Based Cooperation: Reputation-Based Partner Selection and Network}.
\newblock \bibinfo{journal}{\emph{New Frontiers in the Study of Social Phenomena}} (\bibinfo{year}{[n.\,d.]}), \bibinfo{pages}{101}.
\newblock


\bibitem[Vinitsky et~al\mbox{.}(2019)]%
        {SSDOpenSource}
\bibfield{author}{\bibinfo{person}{Eugene Vinitsky}, \bibinfo{person}{Natasha Jaques}, \bibinfo{person}{Joel Leibo}, \bibinfo{person}{Antonio Castenada}, {and} \bibinfo{person}{Edward Hughes}.} \bibinfo{year}{2019}\natexlab{}.
\newblock \showarticletitle{An open source implementation of sequential social dilemma games}.
\newblock \bibinfo{journal}{\emph{Github repository}} (\bibinfo{year}{2019}).
\newblock


\bibitem[Wang et~al\mbox{.}(2022)]%
        {Wang_22}
\bibfield{author}{\bibinfo{person}{Jianrui Wang}, \bibinfo{person}{Yitian Hong}, \bibinfo{person}{Jiali Wang}, \bibinfo{person}{Jiapeng Xu}, \bibinfo{person}{Yang Tang}, \bibinfo{person}{Qing-Long Han}, {and} \bibinfo{person}{J{\"u}rgen Kurths}.} \bibinfo{year}{2022}\natexlab{}.
\newblock \showarticletitle{Cooperative and competitive multi-agent systems: From optimization to games}.
\newblock \bibinfo{journal}{\emph{IEEE/CAA Journal of Automatica Sinica}} \bibinfo{volume}{9}, \bibinfo{number}{5} (\bibinfo{year}{2022}), \bibinfo{pages}{763--783}.
\newblock


\bibitem[Wu et~al\mbox{.}(2016)]%
        {wu_16}
\bibfield{author}{\bibinfo{person}{Junhui Wu}, \bibinfo{person}{Daniel Balliet}, {and} \bibinfo{person}{Paul~AM Van~Lange}.} \bibinfo{year}{2016}\natexlab{}.
\newblock \showarticletitle{Gossip versus punishment: The efficiency of reputation to promote and maintain cooperation}.
\newblock \bibinfo{journal}{\emph{Scientific reports}} \bibinfo{volume}{6}, \bibinfo{number}{1} (\bibinfo{year}{2016}), \bibinfo{pages}{23919}.
\newblock


\end{thebibliography}

%%%%%%%%%%%%%%%%%%%%%%%%%%%%%%%%%%%%%%%%%%%%%%%%%%%%%%%%%%%%%%%%%%%%%%%%
\appendix

\section{Code Availability}
The source code is available at \url{https://anonymous.4open.science/r/Gossip_Agent-7321/}.

\section{Training Details}
\subsection{Prisoner's Dilemma}
The action space for each agent has a size of \(2\): cooperate, defect.

Table \ref{tab:pd_training_detail} presents hyperparameters used in training harvest agents. Default neural networks have two hidden layers with \(64\) units, interleaved with rectified linear layers, and project to an output layer with 2 units, one for each action.

\subsection{Harvest}
The action space for each agent has a size of \(8\): move left, move right, move up, move down, stay, turn clockwise, turn counter-clockwise, and fire beam.

Table \ref{tab:harvest_training_detail} presents hyperparameters used in training harvest agents. Default neural networks have a convolution layer of \(32\) filters with a $3 \times 3$ kernel and a stride of \(1\), followed by two hidden layers with \(32\) and \(16\) units, interleaved with rectified linear layers which projected to the output layer with 8 units, one for each action.
\begin{table}[t]
\centering
\begin{tabular}{clc}
\hline
Description & Value \\
\hline
Num episodes & $2000$ \\
Num steps per episode & $1000$\\
$\gamma$              & $0.99$ \\
Batch size & $128$\\
Learning rate & $1e{-3}$\\

\hline
\end{tabular}
\caption{Hyperparameters for PD trainers.}
\label{tab:pd_training_detail}
\end{table}

\begin{table}[t]
\centering
\begin{tabular}{clc}
\hline
Description & Value \\
\hline
Num episodes & $1e4$ \\
Num steps per episode & $1000$\\
$\gamma$              & $0.99$ \\
Batch size & $100$\\
Learning rate & $1e{-4}$\\

\hline
\end{tabular}
\caption{Hyperparameters for Harvest trainers.}
\label{tab:harvest_training_detail}
\end{table}

\section{Additional Experiments}
\subsection{Ablation Study on Harvest Reputation Rules}
\begin{figure}
    \centering
    \includegraphics[width=\linewidth]{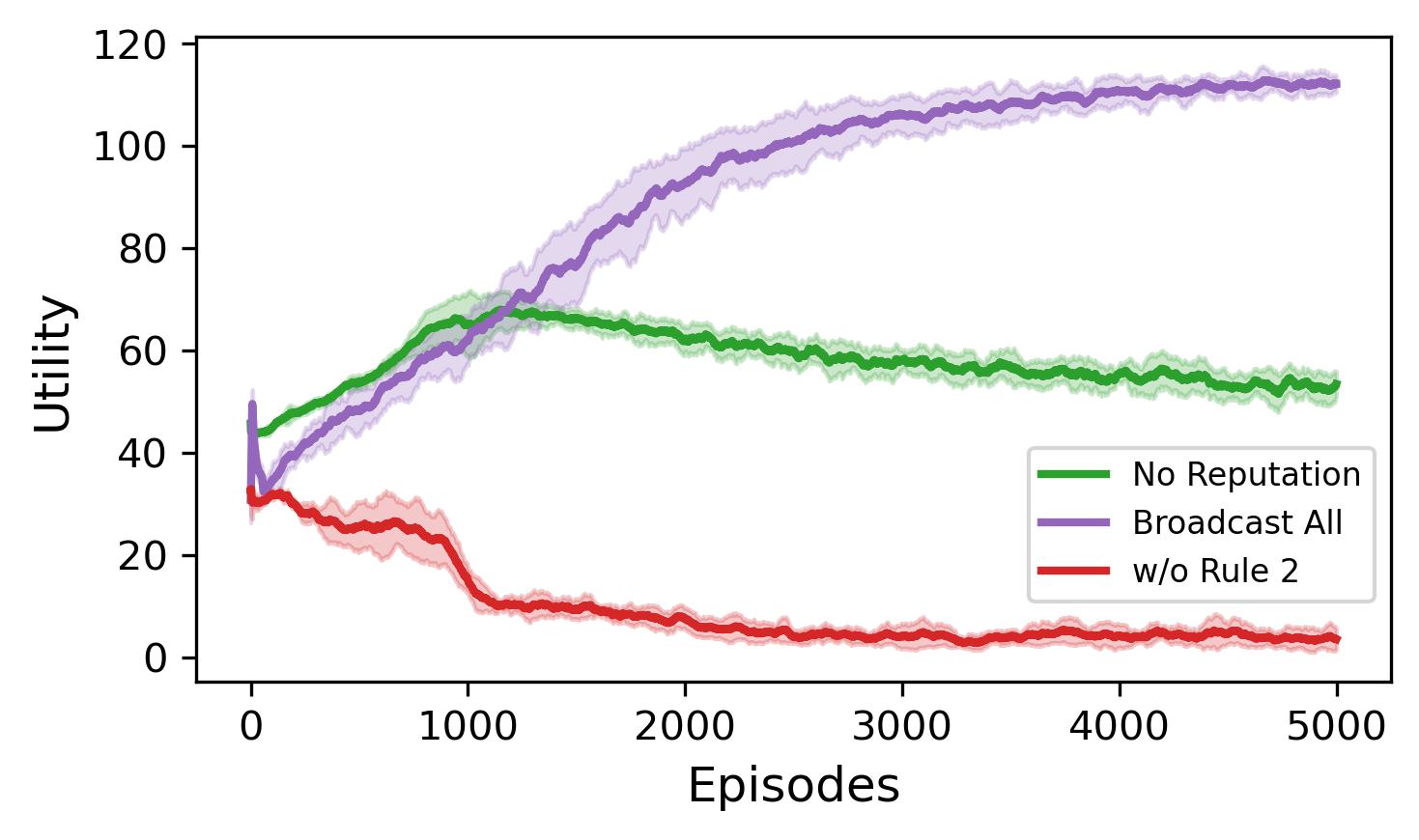}
    \caption{Ablation study performed without }
    \label{fig:ablation}
\end{figure}

Figure \ref{fig:ablation} shows an ablation study on rule 2 of the harvest environment. We set $\rho = \infty$, each curve reports the mean across 5 independent runs, with shaded regions indicating the variance across runs.

\end{document}